\documentclass[reprint, aps, pra, floatfix, superscriptaddress,nofootinbib]{revtex4-1}
\usepackage{graphicx}    % needed for figures
\usepackage{dcolumn}    % needed for some tables
\usepackage{bm}             % for math
\usepackage{amssymb}   % for math
\usepackage{amsmath}    % for multiline equation
\usepackage{amsthm}
\usepackage{mathrsfs}    % for mathscr{}
\usepackage{commath}   % for absolute value
\usepackage{subfigure}    % for subfigure
\usepackage{braket}
\usepackage{natbib}
\usepackage{float}
\usepackage{color}
\usepackage{siunitx}
\usepackage[colorlinks=true, allcolors=blue]{hyperref}
\usepackage{hyphenat}
\graphicspath{{Figures/}}

\newcommand{\beq}{\begin{equation}}
\newcommand{\eeq}{\end{equation}}

\begin{document}

\title{Entanglement of Microwave--Optical Modes in a Strongly Coupled Electro-Optomechanical System}

\author{Changchun Zhong}
\email{zhong.changchun@uchicago.edu}
\affiliation{Pritzker School of Molecular Engineering, University of Chicago, Chicago, IL 60637, USA}
\affiliation{Yale Quantum Institute, Yale University, New Haven, CT 06520, USA}

\author{Xu Han}
\affiliation{Center for Nanoscale Materials, Argonne National Laboratory, Argonne, IL 60439, USA}
\affiliation{Department of Electrical Engineering, Yale University, New Haven, CT 06520, USA}

\author{Hong X. Tang}
\affiliation{Yale Quantum Institute, Yale University, New Haven, CT 06520, USA}
\affiliation{Department of Electrical Engineering, Yale University, New Haven, CT 06520, USA}

\author{Liang Jiang}
\email{liang.jiang@uchicago.edu}
\affiliation{Pritzker School of Molecular Engineering, University of Chicago, Chicago, IL 60637, USA}
\affiliation{Yale Quantum Institute, Yale University, New Haven, CT 06520, USA}

\date{\today}

\begin{abstract}

Quantum transduction between microwave and optics can be realized by quantum teleportation if given reliable microwave-optical entanglement, namely entanglement-based quantum transduction. To realize this protocol, an entangled source with high-fidelity between the two frequencies is necessary. In this paper, we study the microwave and optical entanglement generation based on a generic cavity electro-optomechanical system in the strong coupling regime. Splittings are shown in the microwave and optical output spectra and the frequency entanglement between the two modes is quantified. We show that entanglement can be straightforwardly encoded in the frequency-bin degree of freedom and propose a feasible experiment to verify entangled photon pairs. The experimental implementation is systematically analyzed, and the preferable parameter regime for entanglement verification is identified. An inequality is given as a criterion for good entanglement verification with analysis of practical imperfections.

\end{abstract}

\maketitle

\section{Introduction}

Building a distributed quantum architecture, where distant quantum circuits are connected through low loss optical communication channels, is a long-pursued goal in the quantum computation community \cite{Cirac1997,Kimble2008,Jiang2007,Monroe2014}. To realize this goal, an essential part is to coherently transfer quantum states between the optical channels and the quantum circuits, which in general are in quite different frequency ranges, e.g., optical telecom photons at $\sim200$ THz and superconducting circuits at $\sim10$ GHz microwave frequencies. However, a superconducting qubit doesn't directly interact with optical photons, a high-fidelity quantum transducer is thus in urgent need to interface microwave and optical (M--O) photons in a coherent way. The development of an efficient transducer will not only greatly expand the superconducting quantum network but also connecting superconducting qubits with different quantum modules \cite{Matsukevich2004,Blinov2004,Togan2010,Koehl2011,Gao2012,DeGreve2012,Changhao19}. 

Quantum state transduction can be realized by either direct quantum transduction (DT) which linearly converts photons between different frequencies \cite{Hafezi2012,Gard2017,Hisatomi2016,Tsang2010,Javerzac-Galy2016,Fan2018,Williamson2014,OBrien2014,Andrews2014,Regal2011,Bochmann2013,Taylor2011,Barzanjeh2011,Wang2012,Tian2010,Midolo2018,Bagci2014,Winger2011,Pitanti2015}, or entanglement-based transduction (ET) which first generates entangled photon pairs (or continuous Bosonic modes) with different frequencies then completes the transduction with quantum teleportation \cite{Zhong19,Barzanjeh2012,rueda2019}. {Recently, theoretical proposals given an imperfect DT transducer show the possibility of achieving state transduction by choosing squeezed ancillary input and performing feedforward \cite{Rakhubovsky16,Zhang18,Hoi19}. Experimentally, the feedforward scheme has already shown a great enhancement of the transducer performance \cite{Higginbotham2018}. }Despite these encouraging progress, an efficient quantum-enabled M--O transducer remains to be demonstrated, mainly due to demanding requirements of high conversion efficiency threshold and low added thermal noises. In contrast, ET doesn't require the threshold conversion efficiency, and thus is more compatible with the current technological developments \cite{Vainsencher16,Han2014,Zhong19}. 

A major step in ET is to demonstrate useful entanglement between microwave and optical photon pairs (or continuous modes). In Ref. \cite{Zhong19}, the M--O time-bin entanglement generation and detection based on a generic cavity electro-optomechanical system have been investigated in the weak coupling regime, where a wide range of feasible parameters in this regime can be used to demonstrate M--O entanglement. Especially, the verification could tolerate certain amount of thermal noises, which is compatible with recent experiments: the design of mechanical mode in contact with a $1$ Kelvin thermal bath (to enhance the power handling capability \cite{Zhong19,Mingrui2019}), which is shown to be below the noise threshold. 

In this paper, we propose an M--O frequency-bin entanglement generation and detection scheme based on {a generic electro-optomechanical system} in the strong coupling regime. {For analysis, we consider a cavity piezo-optomechanical system \cite{Han2014,Zou2016}. By exploiting the strong-coupling induced hybridization between the microwave and the mechanical modes, we discuss the frequency-entangled M--O states under an optomechanical parametric down-conversion process.} The entanglement is characterized by calculating the entanglement of formation ($E_\text{F}$) of the output modes. Furthermore, we define an entanglement rate ($E_\text{R}$) to quantify the overall efficiency of entanglement generation, which reaches maximum when the system approaches the exceptional point---a well-studied concept in non-hermitian quantum mechanics \cite{Moiseyev11}. To observe the entanglement experimentally, we propose a heralded scheme that detects the entangled photon pairs in the frequency-bin degree of freedom. We map out the preferable parameter regime satisfying the entanglement criteria (Bell inequality violation or Bell state fidelity). Moreover, the entangled M--O mode correlation function and coincidence count rate of output photons are theoretically estimated. A criterion for good entanglement verification taking into account of dark counts, transmission loss, and detection inefficiency is derived in the end. {The entanglement analysis and proposed detection scheme could be generalized to quantum transducers based on different physical platforms, thus provide a useful framework for analyzing M--O entanglement in the strong coupling regime.}

\section{Piezo-optomechanical system with a blue-detuned drive}
\begin{figure}[t]
\includegraphics[width=\columnwidth]{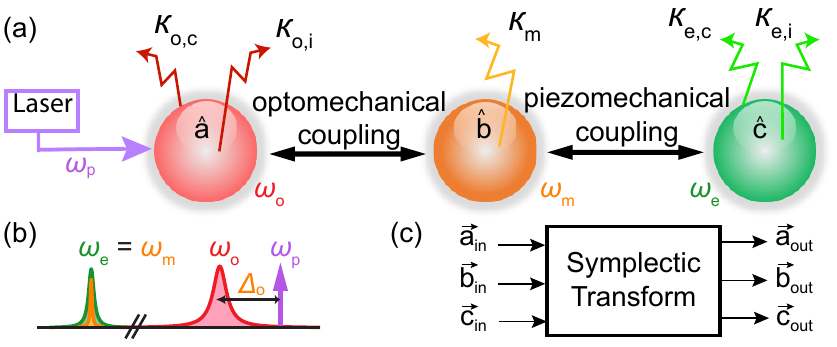}
\caption{(a) Schematic figure for a piezo-optomechanical system with a blue-detuned laser drive. (b) The frequency landscape for microwave, mechanical, optical resonators and the pump laser. (c) The Gaussian unitary transformation connecting input and output mode operators. Take $\vec{\text{a}}_\text{in}=(\hat{{a}}_\text{in,c},\hat{{a}}_\text{in,i})$, and other vectors are similarly defined. \label{fig1}}
\end{figure}

Without losing generality, our discussion is based on a piezo-optomechanical (POM) system with a blue-detuned laser pump, as shown schematically in Fig.~(\ref{fig1}), a mechanical resonator is on one side parametrically coupled to an optical cavity by radiation pressure, and on the other side linearly coupled to a microwave resonator through piezoelectric force. Denoting $\hat{a}$, $\hat{b}$ and $\hat{c}$ as the optical, mechanical, and microwave mode operators, respectively, and $\omega_{\mathrm{o}}$, $\omega_{\mathrm{m}}$, and $\omega_{\mathrm{e}}$ as the corresponding resonant frequencies. Taking $\omega_{\mathrm{e}}=\omega_\mathrm{m}$, we can write down the linearized Hamiltonian of the system with rotating wave approximation \cite{Andrews2014,Zhong19}
\beq\label{hamil}
\begin{split}
\hat{H}=&-\hbar\Delta_\mathrm{o}\hat{a}^\dagger\hat{a}+\hbar\omega_\text{m}{b}^\dagger\hat{b}+\hbar\omega_\text{e}\hat{c}^\dagger\hat{c}\\
&-\hbar g_\text{om}(\hat{a}^\dagger\hat{b}^\dagger+\hat{a}\hat{b})-\hbar g_\text{em}(\hat{b}^\dagger\hat{c}+\hat{b}\hat{c}^\dagger).
\end{split}
\eeq
 {It is worth pointing out that the above Hamiltonian is general for different physical platforms, and thus the theory framework developed below is applicable to various systems \cite{OBrien2014,Andrews2014,Hisatomi2016,Han2014}.} For POM system, $g_\text{em}$ is the piezoelectrical coupling. $g_\text{om}:=\sqrt{\bar{n}_\text{o}}g_\text{om,0}$ is the optomechanical coupling strength, where $g_\text{om,0}$ denotes the single photon coupling. In experiment, the optical cavity will be pumped on the blue sideband by a laser with frequency $\omega_\mathrm{p}=\omega_\mathrm{o}+\Delta_\mathrm{o}$ and is populated with $\bar{n}_\text{o}$ photons on average, which can further enhance the optomechanical coupling. In the discussions that follow, we take the resonance condition $\Delta_\text{o}=\omega_\text{e,m}$.

\begin{table}[t]
\caption{The parameters are used in the numerical evaluations in the text (unless specified otherwise). To comply with the experiment, we leave $C_\text{om}$ and $\kappa_\text{e,c}$ tunable, which can be realized by controlling the optical pump strength and the position of the microwave readout probe.} \label{tab1}
\begin{center}
\begin{tabular}{c|c|c|c|c}
\hline
\hline
$g_{\text{em}}$/MHz  &  $\kappa_ {\text{e,i}}$/kHz & $\kappa_{\text{o,i}}$/(GHz)   & $\kappa_{\text{o,c}}$  & $\kappa_{\text{m}}$/kHz  \\
\hline
$2\pi\times2.0$   & $2\pi\times 100$   &   $2\pi\times 0.24$   & $\kappa_{\text{o,i}}$  &  $2\pi\times20$    \\
\hline
\hline
\end{tabular}
\end{center}
\label{default}
\end{table}

The above system is able to generate entanglement between microwave and optical modes, which is realized intuitively by first entangling the optical and mechanical modes with a two mode squeezing interaction, meanwhile the mechanical excitation is swapped to the microwave mode by the beam-splitter type coupling. To analyze the complete dynamics of the system, we write down the linearized Heisenberg-Langevin equations of motion for each mode,
\beq
\begin{split}\label{eqdy}
\dot{\hat{a}}^\dagger&=(-i\Delta_\text{o}-\frac{\kappa_\text{o}}{2})\hat{a}^\dagger-ig_\text{om}\hat{b}+\sqrt{\kappa_\text{o,c}}\hat{a}^\dagger_\text{in,c}+\sqrt{\kappa_\text{o,i}}\hat{a}^\dagger_\text{in,i},\\
\dot{\hat{b}}&=(-i\omega_\text{m}-\frac{\kappa_\text{m}}{2})\hat{b}+ig_\text{om}\hat{a}^\dagger+ig_\text{em}\hat{c}+\sqrt{\kappa_\text{m}}\hat{b}_\text{in},\\
\dot{\hat{c}}&=(-i\omega_\text{e}-\frac{\kappa_\text{e}}{2})\hat{c}+ig_\text{em}\hat{b}+\sqrt{\kappa_\text{e,c}}\hat{c}_\text{in,c}+\sqrt{\kappa_\text{e,i}}\hat{c}_\text{in,i},
\end{split}
\eeq
where we label the optical, mechanical and microwave decay rates by $\kappa_\text{o}=\kappa_\text{o,c}+\kappa_\text{o,i}$, $\kappa_\text{m}$, and $\kappa_\text{e}=\kappa_\text{e,c}+\kappa_\text{e,i}$. The subscripts ``i" is short for internal loss port, ``c" for coupling port, and ``in" for input noise operator. Equations (\ref{eqdy}) admit a set of hermitian conjugate equations, having essentially the same physics. All input noise operators satisfy \cite{Clerk10}
\beq
\begin{split}
[\hat{o}^\dagger_\text{in}(t),\hat{o}_\text{in}(t^\prime)]&=\bar{n}\delta(t-t^\prime),\\
[\hat{o}_\text{in}(t),\hat{o}^\dagger_\text{in}(t^\prime)]&=(\bar{n}+1)\delta(t-t^\prime).\\
\end{split}
\eeq 
To comply with the experimental condition, we assume the optical resonator and the microwave coupling port are subject to purely vacuum fluctuations $\bar{n}=0$, while the mechanical resonator and the microwave internal port couple to a thermal bath $\bar{n}=\bar{n}_\text{ba}=(e^{\hbar\omega_\text{m(e)}/k_\text{B}T}-1)^{-1}$. The output modes can be obtained by combining the coupled Eq. (\ref{eqdy}) with the input-output formalism (taking optical mode for example)
\beq
\begin{split}
\hat{a}_\text{out,c}&=\sqrt{\kappa_\text{o,c}}\hat{a}-\hat{a}_\text{in,c},\\
\hat{a}_\text{out,i}&=\sqrt{\kappa_\text{o,i}}\hat{a}-\hat{a}_\text{in,i},
\end{split}
\eeq
where the subscript ``out" denotes the output mode. Thus, the system defines a Gaussian unitary channel which is captured by a symplectic transform $\textbf{x}_\text{out}=\textbf{S}\textbf{x}_\text{in}$ \cite{Weedbrook2012,Gosson06}, where $\textbf{S}$ is the symplectic transformation matrix. $\textbf{x}_\text{in}$ and $\textbf{x}_\text{out}$ collect all the input and output quadratures. If we label the M--O output state quadratures as $\textbf{x}=(\hat{x}_\text{o},\hat{p}_\text{o},\hat{x}_\text{e},\hat{p}_\text{e})$, a covariance matrix $\mathbf{V}_{\mathrm{oe}}^\mathrm{out}$ with the elements defined by $V_{ij}=\frac{1}{2} \left\langle \{\hat{x}_i-\braket{\hat{x}_i},\hat{x}_j-\braket{\hat{x}_j} \} \right \rangle$ can be obtained, and it can be expressed in the standard form
\beq\label{vou}
\mathbf{V}_{\mathrm{oe}}^\mathrm{out}
=
\begin{pmatrix}
\mathbf{V}_u & \mathbf{V}_w \\
\mathbf{V}_w & \mathbf{V}_v
\end{pmatrix}
=
\begin{pmatrix}
u(\omega) & 0 & -w(\omega) & 0\\
0 & u(\omega) & 0 & w(\omega)\\
-w(\omega) &0 & v(\omega) & 0\\
0 & w(\omega) & 0 & v(\omega)
\end{pmatrix},
\eeq
where $\textbf{V}_u$, $\textbf{V}_v$ and $\textbf{V}_\text{w}$ are the corresponding two dimensional matrix blocks. This matrix fully characterizes the output M--O Gaussian state, where the diagonal elements represent the corresponding output power spectrum densities and other elements indicate the quadrature correlations.

\section{The piezomechanical strong coupling regime}

\begin{figure}[t]
\includegraphics[width=\columnwidth]{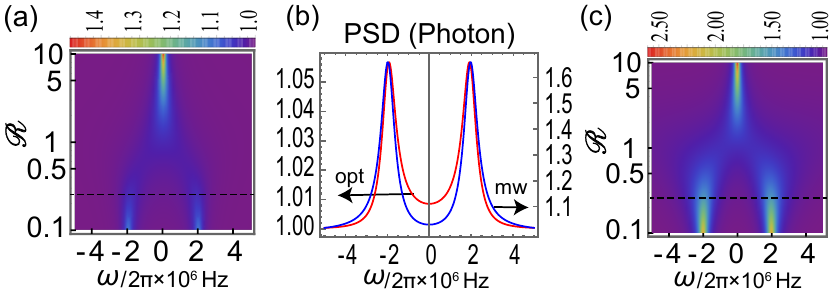}
\caption{  {The output power spectrum densities for (a) optical mode and (c) microwave mode, where mode splitting of $2g_\text{em}$ appears while decreasing the ratio $\mathscr{R}=\kappa_\text{e}/4g_\text{em}$. The ratio $\mathscr{R}=1$ separates the weak and the strong coupling regimes. The dashed lines in (a,c) correspond to the red and blue curves in (b) with the ratio $\mathscr{R}=0.2$.} In these plots, parameters in Tab. \ref{tab1} are used and $\bar{n}_\text{ba}=1$, $C_\text{om}=1$. \label{fig2}}
\end{figure}

Using the feasible parameters given in Tab.~\ref{tab1}, we numerically calculate the M--O output power spectrum densities  {with respect to the ratio $\mathscr{R}=\kappa_\text{e}/4g_\text{em}$, which measures how much the system is strongly ($\mathscr{R}<1$) or weakly ($\mathscr{R}>1$) coupled (details shown later).} As shown in Fig.~(\ref{fig2}), two peaks can be clearly resolved in both the optical and microwave output spectrum as the system approaches the piezomechanical strong coupling regime. Intuitively, the splitting of the microwave output mode is due to the hybridization between the microwave and the mechanical modes while the splitting in the optical mode results from the energy conservation. This can be seen more rigorously from the Eq. (\ref{eqdy}). To show that,
we recast Eq. (\ref{eqdy}) into a more compact form in the microwave rotating frame (we use this frame in all later discussions)
\beq\label{mady}
\dot{\textbf{a}}=\textbf{M}\textbf{a}+\textbf{N}\textbf{a}_\text{in},
\eeq
where we group the operators into the following vectors $\textbf{a}_\text{in}=(\hat{a}^\dagger_\text{in,c},\hat{a}^\dagger_\text{in,i},\hat{b}_\text{in},\hat{c}_\text{in,c},\hat{c}_\text{in,i})^\text{T}$, $\textbf{a}=(\hat{a}^\dagger,\hat{b},\hat{c})^\text{T}$. The matrix
\beq
\textbf{M}=
\begin{pmatrix}
-\frac{\kappa_\text{o}}{2} & -ig_\text{om} & 0\\
ig_\text{om} & -\frac{\kappa_\text{m}}{2} & ig_\text{em}\\
0 & ig_\text{em} & -\frac{\kappa_\text{e}}{2}
\end{pmatrix},
\eeq
\beq
\textbf{N}=
\begin{pmatrix}
\sqrt{\kappa_\text{o,c}} & \sqrt{\kappa_\text{o,i}} &0 &0& 0\\
0&0&\sqrt{\kappa_\text{m}}&0&0\\
0&0&0&\sqrt{\kappa_\text{e,c}}&\sqrt{\kappa_\text{e,i}}
\end{pmatrix}.
\eeq
The non-hermitian dynamical matrix $\textbf{M}$ determines the normal modes. Taking the approximation $\kappa_\text{o}\gg g_\text{om}$ (a relatively lossy optical cavity in experiments), one finds the hybridized normal modes of microwave and mechanical resonator with eigenvalues
\beq
\lambda_{B,C}=-\frac{\kappa_\text{e}+\kappa_\text{m}}{4}\mp\sqrt{ -g_\text{em}^2+\left(\frac{\kappa_\text{e}-\kappa_\text{m}}{4}\right)^2},
\eeq
where the subscripts $B,C$ represent the two hybridized modes. When $g_\text{em}>\abs{\kappa_\text{e}-\kappa_\text{m}}/4$, a negative value in the square root is achieved which corresponds to a mode splitting $2\sqrt{g_\text{em}^2-((\kappa_\text{e}-\kappa_\text{m})/4)^2}$. When $g_\text{em}$ dominates $g_\text{em}\gg\abs{\kappa_\text{e}-\kappa_\text{m}}/4$, the mode splitting approaches $2g_\text{em}\sim2\pi\times4$ MHz, which is exactly what we just showed in Fig.~(\ref{fig2}) in the strong coupling limit. In this limit, we can approximately define the two hybridized modes as \cite{Nonh}
\beq
\begin{split}
\hat{B}=\frac{\sqrt{2}}{2}(\hat{b}+\hat{c}),\\
\hat{C}=\frac{\sqrt{2}}{2}(\hat{b}-\hat{c}),
\end{split}
\eeq
by which the Hamiltonian Eq.~(\ref{hamil}) can be rewritten as
\beq\label{nhami}
\begin{split}
\hat{H}=&-\hbar\Delta_\mathrm{o}\hat{a}^\dagger\hat{a}+\hbar \omega_\text{B}{B}^\dagger\hat{B}+\hbar \omega_\text{C}\hat{C}^\dagger\hat{C}\\
&-\frac{\sqrt{2}\hbar g_\text{om}}{2}(\hat{a}^\dagger\hat{B}^\dagger+\hat{a}\hat{B})-\frac{\sqrt{2}\hbar g_\text{om}}{2}(\hat{a}^\dagger\hat{C}^\dagger+\hat{a}\hat{C}),
\end{split}
\eeq
where $\omega_\text{B}=\omega_\text{m}-g_\text{em}$ and $\omega_\text{C}=\omega_\text{e}+g_\text{em}$ are the new mode frequencies. Thus, we obtain two two-mode-squeezing interactions between the optical mode and each of the hybridized modes, which could simultaneously generate entanglement between either the modes $\hat{a}$ and $\hat{B}$ or $\hat{a}$ and $\hat{C}$. We will show later that this enables us to conveniently encode entangled photon pairs in the frequency degree of freedom. It is worth noting that the above approximation is not always true; however, in our calculation we choose the experimentally compatible parameters with the optomechanical cooperativity $C_\text{om}\sim1$, indicating a good approximation. To be theoretically complete, when $g_\text{em}<\abs{\kappa_\text{e}-\kappa_\text{m}}/4$, a positive value is taken in the square root and the system is called in the weak coupling regime. Especially, when $g_\text{em}=\abs{\kappa_\text{e}-\kappa_\text{m}}/4$, the eigen-values as well as the eigen-vectors coincide, which corresponds to the exceptional point well-known in the non-hermitian quantum physics \cite{Bender98,Moiseyev11,Chen2017}.  {We thus define a ratio $\mathscr{R}=\kappa_\text{e}/4g_\text{em}$ to quantify how much the system is strongly ($\mathscr{R}<1$) or weakly ($\mathscr{R}>1$) coupled, as mentioned before \footnote{   {Strictly speaking, we should define $\mathscr{R}=(\kappa_\text{e}-\kappa_\text{m})/4g_\text{em}$. We left out $\kappa_\text{m}$ since it is relatively small.} }, and $\mathscr{R}=1$ corresponds to the exceptional point. }

\section{Characterization of the output M--O state entanglement}

Ideally, in the strong coupling regime, a product of two-mode squeezed vacuum state can be obtained when analyzing the Hamiltoian Eq.~(\ref{nhami}), written as
\beq
\begin{split}
\ket{\Psi}_\text{eo}\simeq&\sum_{n_1=0}^\infty\frac{r_1^{n_1}}{\sqrt{n_1!}}(\hat{a}_1^\dagger)^{n_1}(\hat{B}^\dagger)^{n_1}\ket{\text{vac}}\\
&\otimes\sum_{n_2=0}^\infty\frac{r_2^{n_2}}{\sqrt{n_2!}}(\hat{a}_2^\dagger)^{n_2}(\hat{C}^\dagger)^{n_2}\ket{\text{vac}},
\end{split}
\eeq
where we define $\hat{a}_1$ and $\hat{a}_2$ as the optical modes that match the frequencies of the hybridized modes $\hat{B}$ and $\hat{C}$ due to energy conservation. $r_1$ and $r_2$ are the effective squeezing factors, which are determined by the optical pump strength and the interaction time before the photons leak out of the cavity. Due to symmetry in our case, we have $r:=r_1=r_2$. For a weak laser pump, $r\ll 1$ and thus
\beq
\ket{\Psi}_\text{eo}\simeq \ket{\text{vac}}+r(\hat{a}_1^\dagger\hat{B}^\dagger+\hat{a}_2^\dagger\hat{C}^\dagger)\ket{\text{vac}}+o(r^2).
\eeq
We see that a Bell state can be generated with probability $\abs{r}^2$. When the state is coupled out of the cavity, neglecting higher order terms in $r$ and discarding the vacuum state (by post selection), we can get a standard Bell state encoded in the frequency-bin degree of freedom.

\begin{figure}[t]
\centering
\includegraphics[width=\columnwidth]{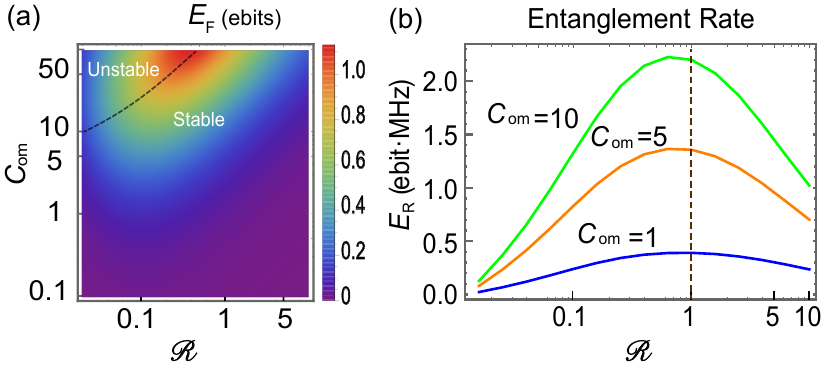}
\caption{  { (a) $E_\text{F}(\omega)$ with $\omega=g_\text{em}=2\pi\times2$ MHz (in rotating frame) in terms of $C_\text{om}$ and the ratio $\mathscr{R}=\kappa_\text{e}/4g_\text{em}$. The dashed line separates the stable and unstable parameter regime. (b) Entanglement rate for varied $C_\text{om}$. The grey vertical line is given by $\mathscr{R}=1$, which separates the weak and the strong coupling regimes.} In these plots, parameters in Tab.~\ref{tab1} are used and $\bar{n}_\text{ba}=1$. \label{fig3}}
\end{figure}

In reality, due to the dissipation and thermal fluctuation, the M--O state is generally a mixed Gaussian state. Thus we can only obtain a mixed two mode Gaussian state, as indicated by the output covariance matrix Eq. (\ref{vou}). In order to characterize the output state, we first use entanglement of formation $E_\text{F}$ to quantify its entanglement. For two mode Gaussian states, $E_\text{F}$ coincides with entanglement cost which quantifies how many Bell states are needed to form a given entangled state. For an output state given in the form Eq.~(\ref{vou}), $E_\text{F}$ can be evaluated by the formula \cite{Marian2008,Tserkis2017}. 
\beq
E_\mathrm{F}(\omega) = \cosh^2r_0 \, \log_2\! \left(\cosh^2 r_0 \right) - \sinh^2r_0 \, \log_2\! \left(\sinh^2 r_0 \right),
\eeq
where $r_0$ is the minimum amount of anti-squeezing needed to disentangled the state and it is given by
\beq
r_0 = \frac{1}{4} \ln \left(\frac{\gamma-\sqrt{\gamma^2-\beta_+\beta_-}}{\beta_-} \right),
\eeq
with
\beq
\begin{split}
\gamma =& 2 \left(\det\mathbf{V}^\mathrm{out}_\mathrm{oe} + 1 \right)-(u(\omega)-v(\omega))^2,\\
\beta_{\pm} =& \det \mathbf{V}_u + \det \mathbf{V}_v-2\det \mathbf{V}_w + 2 u(\omega) v(\omega) \\
&+ 2 w^2(\omega) \pm 4 w(\omega) (u(\omega)+v(\omega)).
\end{split}
\eeq
In general, the squeezing parameter $r_0$ is frequency dependent. For $\omega=0$ and in the low thermal noise limit, it can be simplified to 
\beq
r_0=\frac{1}{2}\ln \frac{1+(\sqrt{C_\text{om}}+\sqrt{C_\text{em}})^2  }{1+(\sqrt{C_\text{om}}-\sqrt{C_\text{em}})^2 },
\eeq
where $C_\text{em}$ is the electromechanical cooperativity. As expected, a bigger squeezing can be obtained when $C_\text{om}\sim C_\text{em}$, which corresponds to the strong parametric down conversion regime. By fixing the output frequency $\omega=2\pi\times2$ MHz \cite{Filt}, we calculate $E_\text{F}$ by scanning the ratio $\mathscr{R}=\kappa_\text{e}/4g_\text{em}$ ($g_\text{em}=2\pi\times 2$ MHz). As shown in Fig. \ref{fig3}(a), entanglement is generated for any non-zero squeezing $r_\text{0}>0$ and reaches maximum along the dashed line, where the system is approaching the strong parametric down conversion regime. Interestingly, this dashed line marks the boundary between the system being stable and not. By numerically checking the stability condition of Eq.~(\ref{mady}) \cite{DeJesus87,Yingdan15,Lin13}, the system is shown to be unstable as $C_\text{om}$ increases, shown by the area above the dashed line in Fig.~\ref{fig3}(a). The reason is that when the blue-detuned laser drive becomes too strong, the optomechanical parametric gain will be too large and cause instability.

The quantity $E_\text{F}(\omega)$ measures the amount of entanglement in the output state for a given frequency. In practice, it is also important to check the entanglement within certain bandwidth.  {Due to energy conservation, the overall output state is approximately in a tensor product form of all frequency contributions, which indicates that the entanglement is additive.} Thus we define a quantity called entanglement rate as
\beq
E_\text{R}=\frac{1}{2\pi}\int E_\text{F}(\omega)d\omega.
\eeq
Intuitively, $E_\text{R}$ tells how efficient a system is in generating entanglement. {In Fig.~\ref{fig3}(b), we calculate $E_\text{R}$ for varied optomechanical cooperativities ($C_\text{om}=1,5,10$) by scanning the ratio $\mathscr{R}=\kappa_\text{e}/4g_\text{em}$ (fixing $g_\text{em}=2\pi\times 2$ MHz) such that the system goes from strong to weak coupling regime. First, we see that the entanglement rate smoothly goes up and down and gives a maximal value when the system is around the exceptional point $\mathscr{R}=1$. The rate $E_\text{R}$ reduces as we further increases or decreases the ratio $\mathscr{R}$. The reason is that a bigger $\mathscr{R}$ means a smaller $C_\text{em}$, leading to further cooperativities mismatch and thus reducing the entanglement rate; while a smaller $\mathscr{R}$ indicates a smaller microwave extraction ratio $\kappa_\text{e,c}/\kappa_\text{e,i}$, effectively decreasing the entanglement rate.} Moreover, Comparing different $C_\text{om}$ in Fig.~\ref{fig3}(b), we find that $E_\text{R}$ is bigger in general for larger $C_\text{om}$ before the system gets unstable, and the peak values of the $E_\text{R}$ shift to the left when increasing $C_\text{om}$, which relates to the fact that the approximation $\kappa_\text{o}\gg g_\text{om}$ is getting worse, such that the exceptional point will shift accordingly.

\section{Verifying Bell state in frequency degree of freedom}

\begin{figure}[t]
\includegraphics[width=\columnwidth]{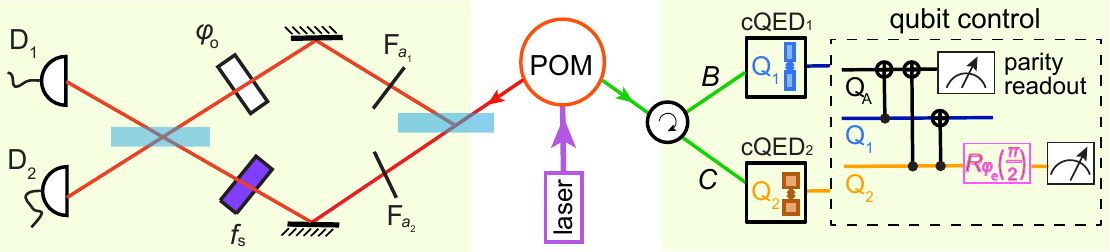}
\caption{Schematic setup for detecting M--O Bell pairs encoded in frequency-bin degree of freedom. The optical photon is analyzed with a balanced Mach-Zehnder interometer, composed of two 50/50 beam splitters (light blue), two narrowband filters (short black line), a phase shifter ($\varphi_\text{o}$) and a frequency shifter $f_\text{s}$. Two single photon detectors are used in the end. The microwave photon detection is realized by cQED systems, where the microwave photon is converted into transmon qubits (Q$_1$ or Q$_2$) excitation by Raman absorption, followed by a partity measurement, a CNOT operation, a $\pi/2$ rotation along the axis ($\sin\varphi_\text{e},\sin\varphi_\text{e},0$) and a high fidelity single qubit readout. \label{fig4}}
\end{figure}

Given the output state feature in the strong coupling regime, we propose an experimental scheme of detecting entangled M--O photon pairs encoded in frequency-bin degree of freedom. As discussed previously, if we decrease the laser pump strength, a standard Bell state in frequency bin is expected in the ideal case \cite{inter}
\beq
\ket{\Psi}_\text{eo}=\frac{\sqrt{2}}{2}(\hat{a}_1^\dagger\hat{B}^\dagger+\hat{a}_2^\dagger\hat{C}^\dagger)\ket{\text{vac}}.
\eeq  
In practice, considering the existence of dissipation and thermal fluctuation, we could only get the output entangled states with certain Bell state fidelity. In this section, we discuss such an experimental scheme to verify the entangled state.

\subsection{The experimental scheme for entanglement verification}

As shown schematically in Fig.~(\ref{fig4}), a POM device driven by a blue detuned laser generates entangled M--O states, whose entanglement property are then detected by the generalized optical and microwave photon detectors shown in the light green blocks. On the optical side, the optical photon is guided into a balanced Mach-Zehnder interferometer with two 50/50 beam splitters. The first beam splitter separates the photon into two paths: one goes through a filter \cite{Narr,Sarah13} which selects $\hat{a}_1$ mode and a phase shifter which shifts a phase $\varphi_\text{o}$; the other goes through a filter selecting $\hat{a}_2$ mode and a frequency shifter \cite{Xianxin17} shifting the mode by frequency $2g_\text{em}$. Then the photons interfere at the second beam splitter. A photon click at the single photon detectors D$_1$ or D$_2$ projects the optical state on $\ket{\varphi_\text{o}}_{\pm}=\frac{\sqrt{2}}{2}(\hat{a}^\dagger_1\pm\hat{a}^\dagger_2e^{i\varphi_\text{o}})\ket{\text{vac}}$. 

On the microwave side, the possible state detection is enabled by two circuit quantum electrodynamical (cQED) systems, each consisting of a transmon qubit with matched dispersive coupling to the cavity modes, respectively.  {In detail, the microwave photon first goes through a circulator, where the modes $\hat{B}$ and $\hat{C}$ are guided into two different cQED systems. The cQED1 and cQED2 are designed to be only resonant with mode $\hat{B}$ and mode $\hat{C}$, respectively, such that mode $\hat{B}$ can only be captured by cQED1 while mode $\hat{C}$ only by cQED2.} The microwave photons are then converted to qubit excitation with the help of stimulated Raman absorption \cite{Campagne-Ibarcq2018}. This step effectively realizes a entanglement swapping from microwave photons to transmon qubits \cite{Eswap}. Immediately after the Raman absorption, a parity measurement is done with the help of an ancillary qubit Q$_\text{A}$ to ensure that one and only one of the two qubits are excited \cite{Nielsen02}. This heralding operation excludes the zero photon and higher order events, increasing significantly the entanglement fidelity. When an odd parity is obtained, we continue performing a CNOT gate to factor out Q$_1$, then apply a $\pi/2$ rotation on Q$_2$ along the axis $(\sin\varphi_\text{e},\cos\varphi_\text{e},0)$ defined on the Bloch sphere. At last, a high fidelity single qubit readout \cite{Shankar2013,Hatridge2013} projects the qubit Q$_2$ onto the state $\ket{\varphi_\text{e}}_\pm=\frac{\sqrt{2}}{2}(\ket{g}\pm\ket{e}e^{-i\varphi_\text{e}})$, which is effectively similar to detecting the microwave state $\ket{\varphi_\text{e}}_\pm=\frac{\sqrt{2}}{2}(\hat{B}^\dagger\pm\hat{C}^\dagger e^{-i\varphi_\text{e}})\ket{\text{vac}}$. 

In summary, the experimental setup allows us to directly measure any states on the equator plane defined on the optical and microwave Bloch spheres. Given two fixed phases $\varphi_\text{o}$ and $\varphi_\text{e}$, repeated measurements could yield the average value
\beq
\begin{split}
E(\varphi_\text{o},\varphi_\text{e})=&p^{+,+}_{\varphi_\text{0},\varphi_\text{e}}+p^{-,-}_{\varphi_\text{0},\varphi_\text{e}}-p^{+,-}_{\varphi_\text{0},\varphi_\text{e}}-p^{-,+}_{\varphi_\text{0},\varphi_\text{e}},
\end{split}
\eeq
where each $p$ denotes the probability of coincident counting clicks for the corresponding state projections, e.g., $p^{+,+}_{\varphi_\text{0},\varphi_\text{e}}=p(\ket{\varphi_\text{0}}_+,\ket{\varphi_\text{e}}_+)$ . Each probability can be theoretically calculated and the details are put in the Appendix.

\subsection{CHSH inequality violation and Bell state fidelity lower bound}

\begin{figure}[t]
\includegraphics[width=\columnwidth]{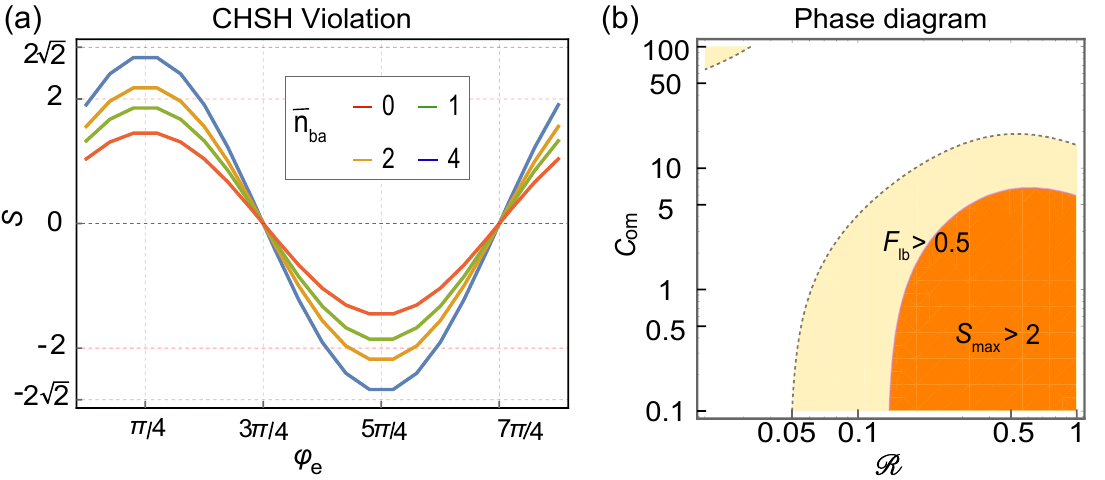}
\caption{(a) The $S$ correlation curves in terms of phase angle $\varphi_\text{e}$ with varied thermal bath, taking $\mathscr{R}=\kappa_\text{e}/4g_\text{em}=0.26$ and $C_\text{om}=1$.  (b) The ``phase diagram'' for the CHSH inequality violation and the Bell state fidelity bigger than $1/2$ in the strong coupling regime with $\mathscr{R}<1$. $\bar{n}_\text{ba}=1$. \label{fig5}}
\end{figure}

The Bell inequality provides a strong manifestation for entanglement, whose violation excludes the possibility of all local hidden variable theory. We use the CHSH-type Bell inequality ($\abs{S}<2$), which can be tested in our proposed experimental setup by measuring the correlation \cite{Clauser1969}
\beq
S=E(\varphi_\text{o},\varphi_\text{e})+E(\varphi_\text{o}^\prime,\varphi_\text{e}^\prime)+E(\varphi_\text{o}^\prime,\varphi_\text{e})-E(\varphi_\text{o},\varphi_\text{e}^\prime).
\eeq
In Fig. \ref{fig5}(a), we simulate the typical correlation curve by fixing $\varphi_\text{o}=0$ and varying the phase $\varphi_\text{e}$ (chose $\varphi_\text{o}^\prime=\varphi_\text{o}+\pi/2,\varphi_\text{e}^\prime=\varphi_\text{e}+\pi/2$). First, a clear Bell inequality violation is observed $\abs{S}>2$ for low thermal noise, indicating the existence of strict entanglement. Also, the violation becomes less obvious as we increase the thermal excitation and the threshold is about two thermal photons. In Fig. \ref{fig5}(b), we map out the parameter regime (in orange) that violates the Bell inequality. We see such regime doesn't overlap the regime where $E_\text{F}$ maximizes, because the regime with maximized $E_\text{F}$ is around the system unstable area, where extremely mixed entangled states is generated that could be unsuitable for Bell test using the proposed experimental setup.

A less demanding evidence for entanglement is given by a Bell state fidelity, which physically measures the closeness between a given state and a standard Bell state. The fidelity being larger than $1/2$ indicates entanglement. In experiment, a quantity easier to measure is the fidelity lower bound, which is given by
\beq
\begin{split}
F_\mathrm{lb} =& \frac{1}{2}( p^{+,+}_{0,0}+p^{-,-}_{0,0}+p^{+,+}_{\frac{\pi}{2},\frac{\pi}{2}}+p^{-,-}_{\frac{\pi}{2},\frac{\pi}{2}}\\
 &- {p^{+,-}_{\frac{\pi}{2},\frac{\pi}{2}}} - {p^{-,+}_{\frac{\pi}{2},\frac{\pi}{2}}} - 2\sqrt{p^{+,-}_{0,0} p^{-,+}_{0,0}} ).
\end{split}
\eeq
In Fig. \ref{fig5}(b), we also delineate the parameter regime where the fidelity lower bound is bigger than $1/2$ (in light yellow). As expected, such regime is much broader than that of CHSH violation since an entangled state is not necessary Bell nonlocal. Due to the reason of easier to be measured, the fidelity lower bound could be a first experiment in the M--O entangled photon pair verifications.

\section{M--O state correlation function and coincidence counting rate}

\begin{figure}[t]
\includegraphics[width=\columnwidth]{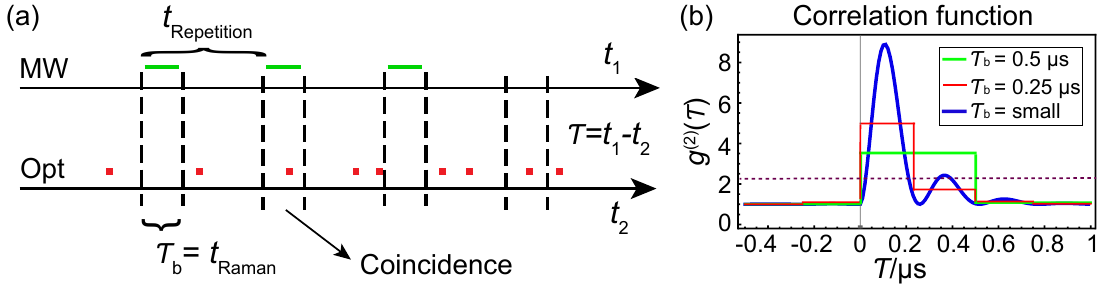}
\caption{(a) Schematic plot for M--O photon coincidence counting measurement. The microwave measurement repetition time is several micro-seconds which includes the qubit preparation, operation and the Raman absorption time that sets the detection time window $\tau_\text{b}$. (b) The second order correlation function for three different detection time resolutions, shown by the blue, red and green lines. For these lines, we have $C_\text{om}=1$, $\kappa_\text{e,c}/\kappa_\text{e,i}=20$ and $\bar{n}_\text{ba}=1$. The dashed horizontal line is determined by $2+\xi_\text{o}+\xi_\text{e}+\xi_\text{o}\xi_\text{e}$. \label{fig6}}
\end{figure}

We start from the well known second order correlation function \cite{Glauber636,*Glauber639}
\beq
g^{(2)}(\tau)=\frac{\braket{\hat{a}_{\text{out,c}}^\dagger(t+\tau)\hat{c}_{\text{out,c}}^\dagger(t)\hat{c}_{\text{out,c}}(t)\hat{a}_{\text{out,c}}(t+\tau) }}{\braket{\hat{a}_{\text{out,c}}^\dagger(t+\tau)\hat{a}_{\text{out,c}}(t+\tau)}\braket{\hat{c}_{\text{out,c}}^\dagger(t)\hat{c}_{\text{out,c}}(t)}},
\eeq
where $\tau$ denotes the time delay between the optical and microwave photon detection. Using the quantum moment-factoring theorem \cite{Shapiro94}, we rewrite the function as
\beq\label{g2f}
g^{(2)}(\tau)=1+\frac{R_\text{oe}(\tau)R_\text{eo}(\tau)}{R_\text{o}R_\text{e}}
\eeq
where $R_\text{o}$ and $R_\text{e}$ are the optical and microwave photon generation rate
\beq
\begin{split}
R_\text{o}=&\braket{\hat{a}_\text{out,c}^\dagger(t)\hat{a}_\text{out,c}(t)}=\frac{1}{2\pi}\int \braket{\hat{a}_\text{out,c}^\dagger({-\omega})\hat{a}_\text{out,c}(-\omega)}d\omega,\\
R_\text{e}=&\braket{\hat{c}_\text{out,c}^\dagger(t)\hat{c}_\text{out,c}(t)}=\frac{1}{2\pi}\int \braket{\hat{c}_\text{out,c}^\dagger({\omega})\hat{c}_\text{out,c}(\omega)}d\omega.
\end{split}
\eeq
$R_\text{oe}(\tau)$ and $R_\text{eo}(\tau)$ are called the M--O correlation rates
\beq
\begin{split}
R_\text{oe}(\tau)=&\braket{\hat{a}_\text{out,c}^\dagger(t+\tau)\hat{c}^\dagger_\text{out,c}(t)}\\
                  =&\frac{1}{2\pi}\int \braket{\hat{a}_\text{out,c}^\dagger({-\omega})\hat{c}^\dagger_\text{out,c}(\omega)}e^{-i\omega\tau}d\omega,\\
R_\text{eo}(\tau)=&\braket{\hat{c}_\text{out,c}(t)\hat{a}_\text{out,c}(t+\tau)}\\
		 =&\frac{1}{2\pi}\int \braket{\hat{c}_\text{out,c}({\omega})\hat{a}_\text{out,c}(-\omega)}e^{i\omega\tau}d\omega.
\end{split}
\eeq 
Equation~(\ref{g2f}) assumes an infinite time resolution of the photon detector. However, as shown in Fig.~\ref{fig6}(a), the photon detector in practice can only resolve photon in a finite time window $\tau_\text{b}$. For the optical detector, the time resolution is generally within nanosecond. While in the microwave detection, the Raman absorption is on the order of microsecond, within which the arrival time of the microwave photon can't be distinguished, and it sets the length of detection time window. Due to this finite time resolution, the measured second order correlation function is generally a piecewise function 
\beq\label{g2m}
\begin{split}
g^{(2)}(\tau_i)&=1+\frac{\int_{\tau_i}^{\tau_i+\tau_\text{b}} R_\text{oe}(\tau)R_\text{eo}(\tau) d\tau}{R_\text{o}R_\text{e}\tau_b},
\end{split}
\eeq
in which $\tau_{i+1}=\tau_i+\tau_\text{b}$. With this formula, we plot the $g^{(2)}(\tau)$ functions with varied detection resolutions in Fig. \ref{fig6}(b). The blue curve gives the ideal case where the detector has infinite time resolution. First, we see an oscillation structure with a period around 0.25 $\mu$s, which is due to frequency beating in the strong coupling regime and it matches exactly with the mode splitting $2g_\text{em}=2\pi\times4$ MHz. Second, the $g^{(2)}(\tau)$ is not symmetric and the maximal value is not happening at the zero time delay. The reason is that the optical and microwave photon have different time profiles, thus their convolution generally is asymmetric. 

\begin{table}[htp]
\caption{Feasible parameters of the photon transmission coefficients, the detector dark count rates and the detector efficiencies.}\label{tab2}
\begin{center}
\begin{tabular}{c|c|c|c|c|c}
\hline
\hline
 $\eta_\text{o}$  & $\eta_\text{e}$ & $D_\text{o}$  & $D_\text{e}$   & $T_e$  & $T_o$  \\
\hline
  $0.8$  & $0.9$   &   $\sim20$ Hz   &  $\sim10^3 $ Hz   & 0.5 & $10^{-3}$  \\
\hline
\hline
\end{tabular}
\end{center}
\end{table}

Equation (\ref{g2m}) can be rewritten as 
\beq
\begin{split}
g^{(2)}(\tau_i)&:=\frac{R_\text{cc}(\tau_i)}{R_\text{ac}}= \frac{R_\text{o}R_\text{e}\tau_\text{b}+\int_{\tau_i}^{\tau_i+\tau_\text{b}} R_\text{oe}(\tau)R_\text{eo}(\tau) d\tau}{R_\text{o}R_\text{e}\tau_\text{b}},
\end{split}
\eeq
where the numerator defines the coincidence counting rate, and we see it contains two parts: the first one is the accidental coincidence rate $R_\text{ac}=R_\text{o}R_\text{e}\tau_b$, while we call the second as the correlated coincidence rate. With the parameters in Tab. \ref{tab1} and take $\tau_\text{b}=0.5\text{ }\mu$s, $C_\text{om}=1$, $\mathscr{R}=0.26$, $\bar{n}_\text{ba}=1$, we find the coincidence counting rate is on the order of $R_\text{cc}\sim10^4$ Hz.

In practice, the experiment also suffers from the photon transmission loss, and detector dark counts and inefficiencies. To give a general model, we denote $T_\text{o}$, $T_\text{e}$ as the optical, microwave transmission coefficients, $D_\text{o}$, $D_\text{e}$ as the optical, microwave detector dark count rates, and $\eta_\text{o}$, $\eta_\text{e}$ as the optical and microwave detector efficiencies, respectively. Taking into account of all these influences and requiring the correlated coincidence rate much larger than the accidental coincidence rate, a simple inequality can be obtained
\beq\label{esi}
g^{(2)}(\tau_i)>2+\xi_\text{o}+\xi_\text{e}+\xi_\text{o}\xi_\text{e},
\eeq
which acts as an useful criterion of ensuring a successful entanglement verification. The quantities $\xi_\text{o}=\frac{D_\text{o}}{\eta_\text{o}T_\text{o}R_\text{o}}$, $\xi_\text{e}=\frac{D_\text{e}}{\eta_\text{e}T_\text{e}R_\text{e}}$, which obviously mean the ratio of dark count rate to photon detection rate. The smaller $\xi_\text{o}$ and $\xi_\text{e}$ are, the better the performance of the experiment will be. With the feasible parameters given in Tab. \ref{tab2}, the coincidence counting rate is reduced to the order of $10$ Hz.  Also, we show the quantity $2+\xi_\text{o}+\xi_\text{e}+\xi_\text{o}\xi_\text{e}\simeq2.37$, which is depicted by a dashed purple line in Fig. \ref{fig6}(b). We see the inequality Eq. (\ref{esi}) can be indeed satisfied as long as the photon detectors have good enough time resolutions. It is worth pointing out that the numerical evaluation is based on current state-of-the-art technological parameters \cite{Han2014}. Given the fast development in this field, we anticipate that the correlated coincidence rate   will be even better. In summary, when designing experiments, all these factors, together with the transmission coefficients and detector efficiencies, must be optimized simultaneously, and the theoretical framework given above provides a useful guide in experimentally manifesting the M--O entanglement.

\section{Discussion}

The entanglement generated from a generic electro-optomechanics can also be investigated in the time-bin degree of freedom, where two short pump pulses with fixed time separation should be applied \cite{Zhong19}. In comparison, this time-bin encoding is suitable for a weakly coupled system, while a strong coupled system is convenient to encode frequency-bin entanglement. Although we can also study frequency-bin entanglement with a weakly coupled system, it generally involves multiple mechanical or microwave modes which would complicate the experimental design. Also, frequency-bin encoding uses a continuous laser drive, which is simpler to implement compare to the time-separated-pulse control in the time-bin encoding. 

Demonstrating the M--O entanglement is the first and important step in DT. Once an entangled source with high-fidelity can be provided, we can then adopt the well-developed teleportation scheme \cite{Bennett93,Bouwmeester97} for quantum transductions. More broadly, the M--O entanglement can also be used to directly entangle distant microwave nodes by adopting the well-known DLCZ scheme \cite{duan2001}. As mentioned in the beginning, ET is compatible with the state-of-the-art technological development, and thus is much less demanding in the experimental implementation. The discussions in this paper thus provide a timely guide for pursuing this direction.

\begin{acknowledgments}
We acknowledge support from the ARL-CDQI (W911NF-15-2-0067, W911NF-18-2-0237), ARO (W911NF-18-1-0020, W911NF-18-1-0212), ARO MURI (W911NF-16-1-0349), AFOSR MURI (FA9550-14-1-0052, FA9550-15-1-0015), DOE (DE-SC0019406), NSF (EFMA-1640959), and the Packard Foundation. This work was performed, in part, at the Center for Nanoscale Materials, a U.S. Department of Energy Office of Science User Facility, and supported by the U.S. Department of Energy, Office of Science, under Contract No. DE-AC02-06CH11357.
\end{acknowledgments}

\begin{appendix}

\section{Coincidence counting probability}

For given output M--O modes with frequency $\omega$, the coincidence counting probability can be theoretically calculated by modeling the optical and microwave detection as 
an on/off photon detector, which is described by a set of positive operator-valued measurements \cite{Ferraro2005}
\begin{align}
\hat{\Pi}_{\mathrm{off}} &= \sum_{n=0}^\infty (1-\eta)^n \ket{n} \bra{n},\\
\hat{\Pi}_{\mathrm{on}} &= \mathbf{I} - \hat{\Pi}_{\mathrm{off}},
\end{align}
in which $\ket{n}$ is photon number state of the mode being detected and $\eta$ models the detector efficiency. Thus, the joint probability is given by $P_\text{on}^\text{o,e}(\omega)=\text{tr}(\hat{\rho}\hat{\Pi}^\text{o}_\text{on}(\omega)\otimes\hat{\Pi}_\text{on}^\text{e}(\omega))$, where $\hat{\rho}$ is the M--O output state density matrix. Since we are dealing with Gaussian states, it is convenient to express $\hat{\rho}$ in the Wigner function
\beq\label{wig}
W(\textbf{x})=\frac{\exp(-\frac{1}{2}\textbf{x}^T\textbf{V}^{-1}\textbf{x})}{(2\pi)^2\sqrt{\text{det}\textbf{V}}},
\eeq
where $\mathbf{V}=\mathbf{V}^\text{out}_\text{oe}(\omega)$ is the corresponding covariance matrix. The coincidence counting probability can be evaluated by
\beq \label{prob_c}
P_\mathrm{on}^\text{o,e}(\omega) = \int W(\mathbf{x}) \tilde{\Pi}_\mathrm{on}^\text{o,e}(\mathbf{x}) \: \mathrm{d} \mathbf{x}.
\eeq
$\tilde{\Pi}_\mathrm{on}^\text{o,e} (\mathbf{x})$ is Weyl transform of $\hat{\Pi}^\text{o}_\mathrm{on}\otimes \hat{\Pi}^\text{e}_\mathrm{on}$ defined by
\beq \label{weyl}
\tilde{\Pi}_\mathrm{on}^\text{o,e} (\mathbf{x}) = \int \left \langle \mathbf{q} + \frac{\mathbf{q}^\prime}{2} \right | \hat{\Pi}^\text{o}_\mathrm{on}\otimes \hat{\Pi}^\text{e}_\mathrm{on} \left | \mathbf{q}-\frac{\mathbf{q}^\prime}{2} \right \rangle e^{i \mathbf{p}^\mathrm{T} \mathbf{q}^\prime} \: \mathrm{d} \mathbf{q}^\prime.
\eeq
where we denote $(\textbf{q},\textbf{p})=(x_\text{o},x_\text{e},p_\text{o},p_\text{e})$. With this formula and taking into account the beam splitter, phase shifter in the experiment, we obtain
\beq\label{coin}
\begin{split}
P_{\varphi_\mathrm{o},\varphi_\mathrm{e}}^{+,+}(\omega)=&1-\frac{2}{(2-\eta_{\mathrm{o}})\sqrt{\mathrm{det}\bm{\Sigma}_{a}}}-\frac{2}{(2-\eta_{\mathrm{e}})\sqrt{\mathrm{det}\bm{\Sigma}_{c}}}\\ 
&+\frac{4}{(2-\eta_{\mathrm{o}})(2-\eta_{\mathrm{e}})\sqrt{\mathrm{det}\bm{\Sigma}_{ac}}}
\end{split}
\eeq
for detecting the states $\ket{\varphi_\mathrm{o}}_+$ and $\ket{\varphi_\mathrm{e}}_+$ simultaneously. The parameters $\eta_\text{o}$ and $\eta_\text{e}$ are the generalized optical and microwave detector efficiencies and
\begin{align}
\mathbf{\Sigma}_{u} &= \frac{\eta_{\mathrm{o}}}{2-\eta_{\mathrm{o}}} \mathbf{V}_{u} + \mathbf{I}_2, \\
\mathbf{\Sigma}_{v} &= \frac{\eta_{\mathrm{e}}}{2-\eta_{\mathrm{e}}} \mathbf{V}_{v} + \mathbf{I}_2, \\
\mathbf{\Sigma}_{w} &= \mathbf{V} \cdot \left( \frac{\eta_{\mathrm{o}}}{2-\eta_{\mathrm{o}}}\mathbf{I}_2 \oplus \frac{\eta_{\mathrm{e}}}{2-\eta_{\mathrm{e}}}\mathbf{I}_2 \right) + \mathbf{I}_4.
\end{align}
Further, the total joint detection rate can be obtained by integrate all frequency contributions,
\beq
\begin{split}
P_{\varphi_\mathrm{o},\varphi_\mathrm{e}}^{+,+} = \int_{\omega_1}^{\omega_2} P_{\varphi_\mathrm{o},\varphi_\mathrm{e}}^{+,+}(\omega) d\omega.
\end{split}
\eeq
The detection rate for other state projections can be derived similarly. In the main text, we use the normalized probability
\beq
p_{\varphi_\mathrm{o},\varphi_\mathrm{e}}^{+,+}=\frac{P_{\varphi_\mathrm{o},\varphi_\mathrm{e}}^{+,+}}{P_{\varphi_\mathrm{o},\varphi_\mathrm{e}}^{+,+}+P_{\varphi_\mathrm{o},\varphi_\mathrm{e}}^{-,-}+P_{\varphi_\mathrm{o},\varphi_\mathrm{e}}^{+,-}+P_{\varphi_\mathrm{o},\varphi_\mathrm{e}}^{-,+}}
\eeq
which corresponds to the physical procedure of post-selecting coincidence counting events based on the heralding signals.

\end{appendix}

\bibliographystyle{apsrev4-1}
\bibliography{Mode_splitting}

\begin{thebibliography}{71}%
\makeatletter
\providecommand \@ifxundefined [1]{%
 \@ifx{#1\undefined}
}%
\providecommand \@ifnum [1]{%
 \ifnum #1\expandafter \@firstoftwo
 \else \expandafter \@secondoftwo
 \fi
}%
\providecommand \@ifx [1]{%
 \ifx #1\expandafter \@firstoftwo
 \else \expandafter \@secondoftwo
 \fi
}%
\providecommand \natexlab [1]{#1}%
\providecommand \enquote  [1]{``#1''}%
\providecommand \bibnamefont  [1]{#1}%
\providecommand \bibfnamefont [1]{#1}%
\providecommand \citenamefont [1]{#1}%
\providecommand \href@noop [0]{\@secondoftwo}%
\providecommand \href [0]{\begingroup \@sanitize@url \@href}%
\providecommand \@href[1]{\@@startlink{#1}\@@href}%
\providecommand \@@href[1]{\endgroup#1\@@endlink}%
\providecommand \@sanitize@url [0]{\catcode `\\12\catcode `\$12\catcode
  `\&12\catcode `\#12\catcode `\^12\catcode `\_12\catcode `\%12\relax}%
\providecommand \@@startlink[1]{}%
\providecommand \@@endlink[0]{}%
\providecommand \url  [0]{\begingroup\@sanitize@url \@url }%
\providecommand \@url [1]{\endgroup\@href {#1}{\urlprefix }}%
\providecommand \urlprefix  [0]{URL }%
\providecommand \Eprint [0]{\href }%
\providecommand \doibase [0]{http://dx.doi.org/}%
\providecommand \selectlanguage [0]{\@gobble}%
\providecommand \bibinfo  [0]{\@secondoftwo}%
\providecommand \bibfield  [0]{\@secondoftwo}%
\providecommand \translation [1]{[#1]}%
\providecommand \BibitemOpen [0]{}%
\providecommand \bibitemStop [0]{}%
\providecommand \bibitemNoStop [0]{.\EOS\space}%
\providecommand \EOS [0]{\spacefactor3000\relax}%
\providecommand \BibitemShut  [1]{\csname bibitem#1\endcsname}%
\let\auto@bib@innerbib\@empty
%</preamble>
\bibitem [{\citenamefont {Cirac}\ \emph {et~al.}(1997)\citenamefont {Cirac},
  \citenamefont {Zoller}, \citenamefont {Kimble},\ and\ \citenamefont
  {Mabuchi}}]{Cirac1997}%
  \BibitemOpen
  \bibfield  {author} {\bibinfo {author} {\bibfnamefont {J.~I.}\ \bibnamefont
  {Cirac}}, \bibinfo {author} {\bibfnamefont {P.}~\bibnamefont {Zoller}},
  \bibinfo {author} {\bibfnamefont {H.~J.}\ \bibnamefont {Kimble}}, \ and\
  \bibinfo {author} {\bibfnamefont {H.}~\bibnamefont {Mabuchi}},\ }\href
  {\doibase 10.1103/PhysRevLett.78.3221} {\bibfield  {journal} {\bibinfo
  {journal} {Phys. Rev. Lett.}\ }\textbf {\bibinfo {volume} {78}},\ \bibinfo
  {pages} {3221} (\bibinfo {year} {1997})}\BibitemShut {NoStop}%
\bibitem [{\citenamefont {Kimble}(2008)}]{Kimble2008}%
  \BibitemOpen
  \bibfield  {author} {\bibinfo {author} {\bibfnamefont {H.~J.}\ \bibnamefont
  {Kimble}},\ }\href {http://dx.doi.org/10.1038/nature07127} {\bibfield
  {journal} {\bibinfo  {journal} {Nature}\ }\textbf {\bibinfo {volume} {453}},\
  \bibinfo {pages} {1023} (\bibinfo {year} {2008})}\BibitemShut {NoStop}%
\bibitem [{\citenamefont {Jiang}\ \emph {et~al.}(2007)\citenamefont {Jiang},
  \citenamefont {Taylor}, \citenamefont {S\o{}rensen},\ and\ \citenamefont
  {Lukin}}]{Jiang2007}%
  \BibitemOpen
  \bibfield  {author} {\bibinfo {author} {\bibfnamefont {L.}~\bibnamefont
  {Jiang}}, \bibinfo {author} {\bibfnamefont {J.~M.}\ \bibnamefont {Taylor}},
  \bibinfo {author} {\bibfnamefont {A.~S.}\ \bibnamefont {S\o{}rensen}}, \ and\
  \bibinfo {author} {\bibfnamefont {M.~D.}\ \bibnamefont {Lukin}},\ }\href
  {\doibase 10.1103/PhysRevA.76.062323} {\bibfield  {journal} {\bibinfo
  {journal} {Phys. Rev. A}\ }\textbf {\bibinfo {volume} {76}},\ \bibinfo
  {pages} {062323} (\bibinfo {year} {2007})}\BibitemShut {NoStop}%
\bibitem [{\citenamefont {Monroe}\ \emph {et~al.}(2014)\citenamefont {Monroe},
  \citenamefont {Raussendorf}, \citenamefont {Ruthven}, \citenamefont {Brown},
  \citenamefont {Maunz}, \citenamefont {Duan},\ and\ \citenamefont
  {Kim}}]{Monroe2014}%
  \BibitemOpen
  \bibfield  {author} {\bibinfo {author} {\bibfnamefont {C.}~\bibnamefont
  {Monroe}}, \bibinfo {author} {\bibfnamefont {R.}~\bibnamefont {Raussendorf}},
  \bibinfo {author} {\bibfnamefont {A.}~\bibnamefont {Ruthven}}, \bibinfo
  {author} {\bibfnamefont {K.~R.}\ \bibnamefont {Brown}}, \bibinfo {author}
  {\bibfnamefont {P.}~\bibnamefont {Maunz}}, \bibinfo {author} {\bibfnamefont
  {L.-M.}\ \bibnamefont {Duan}}, \ and\ \bibinfo {author} {\bibfnamefont
  {J.}~\bibnamefont {Kim}},\ }\href {\doibase 10.1103/PhysRevA.89.022317}
  {\bibfield  {journal} {\bibinfo  {journal} {Phys. Rev. A}\ }\textbf {\bibinfo
  {volume} {89}},\ \bibinfo {pages} {022317} (\bibinfo {year}
  {2014})}\BibitemShut {NoStop}%
\bibitem [{\citenamefont {Matsukevich}\ and\ \citenamefont
  {Kuzmich}(2004)}]{Matsukevich2004}%
  \BibitemOpen
  \bibfield  {author} {\bibinfo {author} {\bibfnamefont {D.~N.}\ \bibnamefont
  {Matsukevich}}\ and\ \bibinfo {author} {\bibfnamefont {A.}~\bibnamefont
  {Kuzmich}},\ }\href {\doibase 10.1126/science.1103346} {\bibfield  {journal}
  {\bibinfo  {journal} {Science}\ }\textbf {\bibinfo {volume} {306}},\ \bibinfo
  {pages} {663} (\bibinfo {year} {2004})}\BibitemShut {NoStop}%
\bibitem [{\citenamefont {Blinov}\ \emph {et~al.}(2004)\citenamefont {Blinov},
  \citenamefont {Moehring}, \citenamefont {Duan},\ and\ \citenamefont
  {Monroe}}]{Blinov2004}%
  \BibitemOpen
  \bibfield  {author} {\bibinfo {author} {\bibfnamefont {B.~B.}\ \bibnamefont
  {Blinov}}, \bibinfo {author} {\bibfnamefont {D.~L.}\ \bibnamefont
  {Moehring}}, \bibinfo {author} {\bibfnamefont {L.-M.}\ \bibnamefont {Duan}},
  \ and\ \bibinfo {author} {\bibfnamefont {C.}~\bibnamefont {Monroe}},\ }\href
  {http://dx.doi.org/10.1038/nature02377} {\bibfield  {journal} {\bibinfo
  {journal} {Nature}\ }\textbf {\bibinfo {volume} {428}},\ \bibinfo {pages}
  {153} (\bibinfo {year} {2004})}\BibitemShut {NoStop}%
\bibitem [{\citenamefont {Togan}\ \emph {et~al.}(2010)\citenamefont {Togan},
  \citenamefont {Chu}, \citenamefont {Trifonov}, \citenamefont {Jiang},
  \citenamefont {Maze}, \citenamefont {Childress}, \citenamefont {Dutt},
  \citenamefont {S{\o}rensen}, \citenamefont {Hemmer}, \citenamefont {Zibrov},\
  and\ \citenamefont {Lukin}}]{Togan2010}%
  \BibitemOpen
  \bibfield  {author} {\bibinfo {author} {\bibfnamefont {E.}~\bibnamefont
  {Togan}}, \bibinfo {author} {\bibfnamefont {Y.}~\bibnamefont {Chu}}, \bibinfo
  {author} {\bibfnamefont {A.~S.}\ \bibnamefont {Trifonov}}, \bibinfo {author}
  {\bibfnamefont {L.}~\bibnamefont {Jiang}}, \bibinfo {author} {\bibfnamefont
  {J.}~\bibnamefont {Maze}}, \bibinfo {author} {\bibfnamefont {L.}~\bibnamefont
  {Childress}}, \bibinfo {author} {\bibfnamefont {M.~V.~G.}\ \bibnamefont
  {Dutt}}, \bibinfo {author} {\bibfnamefont {A.~S.}\ \bibnamefont
  {S{\o}rensen}}, \bibinfo {author} {\bibfnamefont {P.~R.}\ \bibnamefont
  {Hemmer}}, \bibinfo {author} {\bibfnamefont {A.~S.}\ \bibnamefont {Zibrov}},
  \ and\ \bibinfo {author} {\bibfnamefont {M.~D.}\ \bibnamefont {Lukin}},\
  }\href {http://dx.doi.org/10.1038/nature09256} {\bibfield  {journal}
  {\bibinfo  {journal} {Nature}\ }\textbf {\bibinfo {volume} {466}},\ \bibinfo
  {pages} {730} (\bibinfo {year} {2010})}\BibitemShut {NoStop}%
\bibitem [{\citenamefont {Koehl}\ \emph {et~al.}(2011)\citenamefont {Koehl},
  \citenamefont {Buckley}, \citenamefont {Heremans}, \citenamefont {Calusine},\
  and\ \citenamefont {Awschalom}}]{Koehl2011}%
  \BibitemOpen
  \bibfield  {author} {\bibinfo {author} {\bibfnamefont {W.~F.}\ \bibnamefont
  {Koehl}}, \bibinfo {author} {\bibfnamefont {B.~B.}\ \bibnamefont {Buckley}},
  \bibinfo {author} {\bibfnamefont {F.~J.}\ \bibnamefont {Heremans}}, \bibinfo
  {author} {\bibfnamefont {G.}~\bibnamefont {Calusine}}, \ and\ \bibinfo
  {author} {\bibfnamefont {D.~D.}\ \bibnamefont {Awschalom}},\ }\href
  {http://dx.doi.org/10.1038/nature10562} {\bibfield  {journal} {\bibinfo
  {journal} {Nature}\ }\textbf {\bibinfo {volume} {479}},\ \bibinfo {pages}
  {84} (\bibinfo {year} {2011})}\BibitemShut {NoStop}%
\bibitem [{\citenamefont {Gao}\ \emph {et~al.}(2012)\citenamefont {Gao},
  \citenamefont {Fallahi}, \citenamefont {Togan}, \citenamefont
  {Miguel-Sanchez},\ and\ \citenamefont {Imamoglu}}]{Gao2012}%
  \BibitemOpen
  \bibfield  {author} {\bibinfo {author} {\bibfnamefont {W.~B.}\ \bibnamefont
  {Gao}}, \bibinfo {author} {\bibfnamefont {P.}~\bibnamefont {Fallahi}},
  \bibinfo {author} {\bibfnamefont {E.}~\bibnamefont {Togan}}, \bibinfo
  {author} {\bibfnamefont {J.}~\bibnamefont {Miguel-Sanchez}}, \ and\ \bibinfo
  {author} {\bibfnamefont {A.}~\bibnamefont {Imamoglu}},\ }\href
  {http://dx.doi.org/10.1038/nature11573} {\bibfield  {journal} {\bibinfo
  {journal} {Nature}\ }\textbf {\bibinfo {volume} {491}},\ \bibinfo {pages}
  {426} (\bibinfo {year} {2012})}\BibitemShut {NoStop}%
\bibitem [{\citenamefont {De~Greve}\ \emph {et~al.}(2012)\citenamefont
  {De~Greve}, \citenamefont {Yu}, \citenamefont {McMahon}, \citenamefont
  {Pelc}, \citenamefont {Natarajan}, \citenamefont {Kim}, \citenamefont {Abe},
  \citenamefont {Maier}, \citenamefont {Schneider}, \citenamefont {Kamp},
  \citenamefont {H{\"o}fling}, \citenamefont {Hadfield}, \citenamefont
  {Forchel}, \citenamefont {Fejer},\ and\ \citenamefont
  {Yamamoto}}]{DeGreve2012}%
  \BibitemOpen
  \bibfield  {author} {\bibinfo {author} {\bibfnamefont {K.}~\bibnamefont
  {De~Greve}}, \bibinfo {author} {\bibfnamefont {L.}~\bibnamefont {Yu}},
  \bibinfo {author} {\bibfnamefont {P.~L.}\ \bibnamefont {McMahon}}, \bibinfo
  {author} {\bibfnamefont {J.~S.}\ \bibnamefont {Pelc}}, \bibinfo {author}
  {\bibfnamefont {C.~M.}\ \bibnamefont {Natarajan}}, \bibinfo {author}
  {\bibfnamefont {N.~Y.}\ \bibnamefont {Kim}}, \bibinfo {author} {\bibfnamefont
  {E.}~\bibnamefont {Abe}}, \bibinfo {author} {\bibfnamefont {S.}~\bibnamefont
  {Maier}}, \bibinfo {author} {\bibfnamefont {C.}~\bibnamefont {Schneider}},
  \bibinfo {author} {\bibfnamefont {M.}~\bibnamefont {Kamp}}, \bibinfo {author}
  {\bibfnamefont {S.}~\bibnamefont {H{\"o}fling}}, \bibinfo {author}
  {\bibfnamefont {R.~H.}\ \bibnamefont {Hadfield}}, \bibinfo {author}
  {\bibfnamefont {A.}~\bibnamefont {Forchel}}, \bibinfo {author} {\bibfnamefont
  {M.~M.}\ \bibnamefont {Fejer}}, \ and\ \bibinfo {author} {\bibfnamefont
  {Y.}~\bibnamefont {Yamamoto}},\ }\href
  {http://dx.doi.org/10.1038/nature11577} {\bibfield  {journal} {\bibinfo
  {journal} {Nature}\ }\textbf {\bibinfo {volume} {491}},\ \bibinfo {pages}
  {421} (\bibinfo {year} {2012})}\BibitemShut {NoStop}%
\bibitem [{\citenamefont {Li}\ and\ \citenamefont
  {Cappellaro}(2019)}]{Changhao19}%
  \BibitemOpen
  \bibfield  {author} {\bibinfo {author} {\bibfnamefont {C.}~\bibnamefont
  {Li}}\ and\ \bibinfo {author} {\bibfnamefont {P.}~\bibnamefont
  {Cappellaro}},\ }\href@noop {} {\bibfield  {journal} {\bibinfo  {journal}
  {arXiv preprint arXiv:1904.01556}\ } (\bibinfo {year} {2019})}\BibitemShut
  {NoStop}%
\bibitem [{\citenamefont {Hafezi}\ \emph {et~al.}(2012)\citenamefont {Hafezi},
  \citenamefont {Kim}, \citenamefont {Rolston}, \citenamefont {Orozco},
  \citenamefont {Lev},\ and\ \citenamefont {Taylor}}]{Hafezi2012}%
  \BibitemOpen
  \bibfield  {author} {\bibinfo {author} {\bibfnamefont {M.}~\bibnamefont
  {Hafezi}}, \bibinfo {author} {\bibfnamefont {Z.}~\bibnamefont {Kim}},
  \bibinfo {author} {\bibfnamefont {S.~L.}\ \bibnamefont {Rolston}}, \bibinfo
  {author} {\bibfnamefont {L.~A.}\ \bibnamefont {Orozco}}, \bibinfo {author}
  {\bibfnamefont {B.~L.}\ \bibnamefont {Lev}}, \ and\ \bibinfo {author}
  {\bibfnamefont {J.~M.}\ \bibnamefont {Taylor}},\ }\href {\doibase
  10.1103/PhysRevA.85.020302} {\bibfield  {journal} {\bibinfo  {journal} {Phys.
  Rev. A}\ }\textbf {\bibinfo {volume} {85}},\ \bibinfo {pages} {020302(R)}
  (\bibinfo {year} {2012})}\BibitemShut {NoStop}%
\bibitem [{\citenamefont {Gard}\ \emph {et~al.}(2017)\citenamefont {Gard},
  \citenamefont {Jacobs}, \citenamefont {McDermott},\ and\ \citenamefont
  {Saffman}}]{Gard2017}%
  \BibitemOpen
  \bibfield  {author} {\bibinfo {author} {\bibfnamefont {B.~T.}\ \bibnamefont
  {Gard}}, \bibinfo {author} {\bibfnamefont {K.}~\bibnamefont {Jacobs}},
  \bibinfo {author} {\bibfnamefont {R.}~\bibnamefont {McDermott}}, \ and\
  \bibinfo {author} {\bibfnamefont {M.}~\bibnamefont {Saffman}},\ }\href
  {\doibase 10.1103/PhysRevA.96.013833} {\bibfield  {journal} {\bibinfo
  {journal} {Phys. Rev. A}\ }\textbf {\bibinfo {volume} {96}},\ \bibinfo
  {pages} {013833} (\bibinfo {year} {2017})}\BibitemShut {NoStop}%
\bibitem [{\citenamefont {Hisatomi}\ \emph {et~al.}(2016)\citenamefont
  {Hisatomi}, \citenamefont {Osada}, \citenamefont {Tabuchi}, \citenamefont
  {Ishikawa}, \citenamefont {Noguchi}, \citenamefont {Yamazaki}, \citenamefont
  {Usami},\ and\ \citenamefont {Nakamura}}]{Hisatomi2016}%
  \BibitemOpen
  \bibfield  {author} {\bibinfo {author} {\bibfnamefont {R.}~\bibnamefont
  {Hisatomi}}, \bibinfo {author} {\bibfnamefont {A.}~\bibnamefont {Osada}},
  \bibinfo {author} {\bibfnamefont {Y.}~\bibnamefont {Tabuchi}}, \bibinfo
  {author} {\bibfnamefont {T.}~\bibnamefont {Ishikawa}}, \bibinfo {author}
  {\bibfnamefont {A.}~\bibnamefont {Noguchi}}, \bibinfo {author} {\bibfnamefont
  {R.}~\bibnamefont {Yamazaki}}, \bibinfo {author} {\bibfnamefont
  {K.}~\bibnamefont {Usami}}, \ and\ \bibinfo {author} {\bibfnamefont
  {Y.}~\bibnamefont {Nakamura}},\ }\href {\doibase 10.1103/PhysRevB.93.174427}
  {\bibfield  {journal} {\bibinfo  {journal} {Phys. Rev. B}\ }\textbf {\bibinfo
  {volume} {93}},\ \bibinfo {pages} {174427} (\bibinfo {year}
  {2016})}\BibitemShut {NoStop}%
\bibitem [{\citenamefont {Tsang}(2010)}]{Tsang2010}%
  \BibitemOpen
  \bibfield  {author} {\bibinfo {author} {\bibfnamefont {M.}~\bibnamefont
  {Tsang}},\ }\href {\doibase 10.1103/PhysRevA.81.063837} {\bibfield  {journal}
  {\bibinfo  {journal} {Phys. Rev. A}\ }\textbf {\bibinfo {volume} {81}},\
  \bibinfo {pages} {063837} (\bibinfo {year} {2010})}\BibitemShut {NoStop}%
\bibitem [{\citenamefont {Javerzac-Galy}\ \emph {et~al.}(2016)\citenamefont
  {Javerzac-Galy}, \citenamefont {Plekhanov}, \citenamefont {Bernier},
  \citenamefont {Toth}, \citenamefont {Feofanov},\ and\ \citenamefont
  {Kippenberg}}]{Javerzac-Galy2016}%
  \BibitemOpen
  \bibfield  {author} {\bibinfo {author} {\bibfnamefont {C.}~\bibnamefont
  {Javerzac-Galy}}, \bibinfo {author} {\bibfnamefont {K.}~\bibnamefont
  {Plekhanov}}, \bibinfo {author} {\bibfnamefont {N.~R.}\ \bibnamefont
  {Bernier}}, \bibinfo {author} {\bibfnamefont {L.~D.}\ \bibnamefont {Toth}},
  \bibinfo {author} {\bibfnamefont {A.~K.}\ \bibnamefont {Feofanov}}, \ and\
  \bibinfo {author} {\bibfnamefont {T.~J.}\ \bibnamefont {Kippenberg}},\ }\href
  {\doibase 10.1103/PhysRevA.94.053815} {\bibfield  {journal} {\bibinfo
  {journal} {Phys. Rev. A}\ }\textbf {\bibinfo {volume} {94}},\ \bibinfo
  {pages} {053815} (\bibinfo {year} {2016})}\BibitemShut {NoStop}%
\bibitem [{\citenamefont {Fan}\ \emph {et~al.}()\citenamefont {Fan},
  \citenamefont {Zou}, \citenamefont {Cheng}, \citenamefont {Guo},
  \citenamefont {Han}, \citenamefont {Gong}, \citenamefont {Wang},\ and\
  \citenamefont {Tang}}]{Fan2018}%
  \BibitemOpen
  \bibfield  {author} {\bibinfo {author} {\bibfnamefont {L.}~\bibnamefont
  {Fan}}, \bibinfo {author} {\bibfnamefont {C.-L.}\ \bibnamefont {Zou}},
  \bibinfo {author} {\bibfnamefont {R.}~\bibnamefont {Cheng}}, \bibinfo
  {author} {\bibfnamefont {X.}~\bibnamefont {Guo}}, \bibinfo {author}
  {\bibfnamefont {X.}~\bibnamefont {Han}}, \bibinfo {author} {\bibfnamefont
  {Z.}~\bibnamefont {Gong}}, \bibinfo {author} {\bibfnamefont {S.}~\bibnamefont
  {Wang}}, \ and\ \bibinfo {author} {\bibfnamefont {H.~X.}\ \bibnamefont
  {Tang}},\ }\href@noop {} {\ }\bibinfo {note}
  {\href{10.1126/sciadv.aar4994}{Sci. Adv. \text{Vol. 4, no. 8, eaar4994}
  (2018)}}\BibitemShut {NoStop}%
\bibitem [{\citenamefont {Williamson}\ \emph {et~al.}(2014)\citenamefont
  {Williamson}, \citenamefont {Chen},\ and\ \citenamefont
  {Longdell}}]{Williamson2014}%
  \BibitemOpen
  \bibfield  {author} {\bibinfo {author} {\bibfnamefont {L.~A.}\ \bibnamefont
  {Williamson}}, \bibinfo {author} {\bibfnamefont {Y.-H.}\ \bibnamefont
  {Chen}}, \ and\ \bibinfo {author} {\bibfnamefont {J.~J.}\ \bibnamefont
  {Longdell}},\ }\href {\doibase 10.1103/PhysRevLett.113.203601} {\bibfield
  {journal} {\bibinfo  {journal} {Phys. Rev. Lett.}\ }\textbf {\bibinfo
  {volume} {113}},\ \bibinfo {pages} {203601} (\bibinfo {year}
  {2014})}\BibitemShut {NoStop}%
\bibitem [{\citenamefont {O'Brien}\ \emph {et~al.}(2014)\citenamefont
  {O'Brien}, \citenamefont {Lauk}, \citenamefont {Blum}, \citenamefont
  {Morigi},\ and\ \citenamefont {Fleischhauer}}]{OBrien2014}%
  \BibitemOpen
  \bibfield  {author} {\bibinfo {author} {\bibfnamefont {C.}~\bibnamefont
  {O'Brien}}, \bibinfo {author} {\bibfnamefont {N.}~\bibnamefont {Lauk}},
  \bibinfo {author} {\bibfnamefont {S.}~\bibnamefont {Blum}}, \bibinfo {author}
  {\bibfnamefont {G.}~\bibnamefont {Morigi}}, \ and\ \bibinfo {author}
  {\bibfnamefont {M.}~\bibnamefont {Fleischhauer}},\ }\href {\doibase
  10.1103/PhysRevLett.113.063603} {\bibfield  {journal} {\bibinfo  {journal}
  {Phys. Rev. Lett.}\ }\textbf {\bibinfo {volume} {113}},\ \bibinfo {pages}
  {063603} (\bibinfo {year} {2014})}\BibitemShut {NoStop}%
\bibitem [{\citenamefont {Andrews}\ \emph {et~al.}(2014)\citenamefont
  {Andrews}, \citenamefont {Peterson}, \citenamefont {Purdy}, \citenamefont
  {Cicak}, \citenamefont {Simmonds}, \citenamefont {Regal},\ and\ \citenamefont
  {Lehnert}}]{Andrews2014}%
  \BibitemOpen
  \bibfield  {author} {\bibinfo {author} {\bibfnamefont {R.~W.}\ \bibnamefont
  {Andrews}}, \bibinfo {author} {\bibfnamefont {R.~W.}\ \bibnamefont
  {Peterson}}, \bibinfo {author} {\bibfnamefont {T.~P.}\ \bibnamefont {Purdy}},
  \bibinfo {author} {\bibfnamefont {K.}~\bibnamefont {Cicak}}, \bibinfo
  {author} {\bibfnamefont {R.~W.}\ \bibnamefont {Simmonds}}, \bibinfo {author}
  {\bibfnamefont {C.~A.}\ \bibnamefont {Regal}}, \ and\ \bibinfo {author}
  {\bibfnamefont {K.~W.}\ \bibnamefont {Lehnert}},\ }\href
  {http://dx.doi.org/10.1038/nphys2911} {\bibfield  {journal} {\bibinfo
  {journal} {Nat. Phys.}\ }\textbf {\bibinfo {volume} {10}},\ \bibinfo {pages}
  {321} (\bibinfo {year} {2014})}\BibitemShut {NoStop}%
\bibitem [{\citenamefont {Regal}\ and\ \citenamefont
  {Lehnert}(2011)}]{Regal2011}%
  \BibitemOpen
  \bibfield  {author} {\bibinfo {author} {\bibfnamefont {C.~A.}\ \bibnamefont
  {Regal}}\ and\ \bibinfo {author} {\bibfnamefont {K.~W.}\ \bibnamefont
  {Lehnert}},\ }\href {\doibase 10.1088/1742-6596/264/1/012025} {\bibfield
  {journal} {\bibinfo  {journal} {J. Phys.: Conf. Ser.}\ }\textbf {\bibinfo
  {volume} {264}},\ \bibinfo {pages} {012025} (\bibinfo {year}
  {2011})}\BibitemShut {NoStop}%
\bibitem [{\citenamefont {Bochmann}\ \emph {et~al.}(2013)\citenamefont
  {Bochmann}, \citenamefont {Vainsencher}, \citenamefont {Awschalom},\ and\
  \citenamefont {Cleland}}]{Bochmann2013}%
  \BibitemOpen
  \bibfield  {author} {\bibinfo {author} {\bibfnamefont {J.}~\bibnamefont
  {Bochmann}}, \bibinfo {author} {\bibfnamefont {A.}~\bibnamefont
  {Vainsencher}}, \bibinfo {author} {\bibfnamefont {D.~D.}\ \bibnamefont
  {Awschalom}}, \ and\ \bibinfo {author} {\bibfnamefont {A.~N.}\ \bibnamefont
  {Cleland}},\ }\href {http://dx.doi.org/10.1038/nphys2748} {\bibfield
  {journal} {\bibinfo  {journal} {Nat. Phys.}\ }\textbf {\bibinfo {volume}
  {9}},\ \bibinfo {pages} {712} (\bibinfo {year} {2013})}\BibitemShut {NoStop}%
\bibitem [{\citenamefont {Taylor}\ \emph {et~al.}(2011)\citenamefont {Taylor},
  \citenamefont {S\o{}rensen}, \citenamefont {Marcus},\ and\ \citenamefont
  {Polzik}}]{Taylor2011}%
  \BibitemOpen
  \bibfield  {author} {\bibinfo {author} {\bibfnamefont {J.~M.}\ \bibnamefont
  {Taylor}}, \bibinfo {author} {\bibfnamefont {A.~S.}\ \bibnamefont
  {S\o{}rensen}}, \bibinfo {author} {\bibfnamefont {C.~M.}\ \bibnamefont
  {Marcus}}, \ and\ \bibinfo {author} {\bibfnamefont {E.~S.}\ \bibnamefont
  {Polzik}},\ }\href {\doibase 10.1103/PhysRevLett.107.273601} {\bibfield
  {journal} {\bibinfo  {journal} {Phys. Rev. Lett.}\ }\textbf {\bibinfo
  {volume} {107}},\ \bibinfo {pages} {273601} (\bibinfo {year}
  {2011})}\BibitemShut {NoStop}%
\bibitem [{\citenamefont {Barzanjeh}\ \emph {et~al.}(2011)\citenamefont
  {Barzanjeh}, \citenamefont {Vitali}, \citenamefont {Tombesi},\ and\
  \citenamefont {Milburn}}]{Barzanjeh2011}%
  \BibitemOpen
  \bibfield  {author} {\bibinfo {author} {\bibfnamefont {S.}~\bibnamefont
  {Barzanjeh}}, \bibinfo {author} {\bibfnamefont {D.}~\bibnamefont {Vitali}},
  \bibinfo {author} {\bibfnamefont {P.}~\bibnamefont {Tombesi}}, \ and\
  \bibinfo {author} {\bibfnamefont {G.~J.}\ \bibnamefont {Milburn}},\ }\href
  {\doibase 10.1103/PhysRevA.84.042342} {\bibfield  {journal} {\bibinfo
  {journal} {Phys. Rev. A}\ }\textbf {\bibinfo {volume} {84}},\ \bibinfo
  {pages} {042342} (\bibinfo {year} {2011})}\BibitemShut {NoStop}%
\bibitem [{\citenamefont {Wang}\ and\ \citenamefont {Clerk}(2012)}]{Wang2012}%
  \BibitemOpen
  \bibfield  {author} {\bibinfo {author} {\bibfnamefont {Y.-D.}\ \bibnamefont
  {Wang}}\ and\ \bibinfo {author} {\bibfnamefont {A.~A.}\ \bibnamefont
  {Clerk}},\ }\href {\doibase 10.1103/PhysRevLett.108.153603} {\bibfield
  {journal} {\bibinfo  {journal} {Phys. Rev. Lett.}\ }\textbf {\bibinfo
  {volume} {108}},\ \bibinfo {pages} {153603} (\bibinfo {year}
  {2012})}\BibitemShut {NoStop}%
\bibitem [{\citenamefont {Tian}\ and\ \citenamefont {Wang}(2010)}]{Tian2010}%
  \BibitemOpen
  \bibfield  {author} {\bibinfo {author} {\bibfnamefont {L.}~\bibnamefont
  {Tian}}\ and\ \bibinfo {author} {\bibfnamefont {H.}~\bibnamefont {Wang}},\
  }\href {\doibase 10.1103/PhysRevA.82.053806} {\bibfield  {journal} {\bibinfo
  {journal} {Phys. Rev. A}\ }\textbf {\bibinfo {volume} {82}},\ \bibinfo
  {pages} {053806} (\bibinfo {year} {2010})}\BibitemShut {NoStop}%
\bibitem [{\citenamefont {Midolo}\ \emph {et~al.}(2018)\citenamefont {Midolo},
  \citenamefont {Schliesser},\ and\ \citenamefont {Fiore}}]{Midolo2018}%
  \BibitemOpen
  \bibfield  {author} {\bibinfo {author} {\bibfnamefont {L.}~\bibnamefont
  {Midolo}}, \bibinfo {author} {\bibfnamefont {A.}~\bibnamefont {Schliesser}},
  \ and\ \bibinfo {author} {\bibfnamefont {A.}~\bibnamefont {Fiore}},\ }\href
  {\doibase 10.1038/s41565-017-0039-1} {\bibfield  {journal} {\bibinfo
  {journal} {Nat. Nanotech.}\ }\textbf {\bibinfo {volume} {13}},\ \bibinfo
  {pages} {11} (\bibinfo {year} {2018})}\BibitemShut {NoStop}%
\bibitem [{\citenamefont {Bagci}\ \emph {et~al.}(2014)\citenamefont {Bagci},
  \citenamefont {Simonsen}, \citenamefont {Schmid}, \citenamefont {Villanueva},
  \citenamefont {Zeuthen}, \citenamefont {Appel}, \citenamefont {Taylor},
  \citenamefont {S{\o}rensen}, \citenamefont {Usami}, \citenamefont
  {Schliesser},\ and\ \citenamefont {Polzik}}]{Bagci2014}%
  \BibitemOpen
  \bibfield  {author} {\bibinfo {author} {\bibfnamefont {T.}~\bibnamefont
  {Bagci}}, \bibinfo {author} {\bibfnamefont {A.}~\bibnamefont {Simonsen}},
  \bibinfo {author} {\bibfnamefont {S.}~\bibnamefont {Schmid}}, \bibinfo
  {author} {\bibfnamefont {L.~G.}\ \bibnamefont {Villanueva}}, \bibinfo
  {author} {\bibfnamefont {E.}~\bibnamefont {Zeuthen}}, \bibinfo {author}
  {\bibfnamefont {J.}~\bibnamefont {Appel}}, \bibinfo {author} {\bibfnamefont
  {J.~M.}\ \bibnamefont {Taylor}}, \bibinfo {author} {\bibfnamefont
  {A.}~\bibnamefont {S{\o}rensen}}, \bibinfo {author} {\bibfnamefont
  {K.}~\bibnamefont {Usami}}, \bibinfo {author} {\bibfnamefont
  {A.}~\bibnamefont {Schliesser}}, \ and\ \bibinfo {author} {\bibfnamefont
  {E.~S.}\ \bibnamefont {Polzik}},\ }\href
  {http://dx.doi.org/10.1038/nature13029} {\bibfield  {journal} {\bibinfo
  {journal} {Nature}\ }\textbf {\bibinfo {volume} {507}},\ \bibinfo {pages}
  {81} (\bibinfo {year} {2014})}\BibitemShut {NoStop}%
\bibitem [{\citenamefont {Winger}\ \emph {et~al.}(2011)\citenamefont {Winger},
  \citenamefont {Blasius}, \citenamefont {Alegre}, \citenamefont
  {Safavi-Naeini}, \citenamefont {Meenehan}, \citenamefont {Cohen},
  \citenamefont {Stobbe},\ and\ \citenamefont {Painter}}]{Winger2011}%
  \BibitemOpen
  \bibfield  {author} {\bibinfo {author} {\bibfnamefont {M.}~\bibnamefont
  {Winger}}, \bibinfo {author} {\bibfnamefont {T.~D.}\ \bibnamefont {Blasius}},
  \bibinfo {author} {\bibfnamefont {T.~P.~M.}\ \bibnamefont {Alegre}}, \bibinfo
  {author} {\bibfnamefont {A.~H.}\ \bibnamefont {Safavi-Naeini}}, \bibinfo
  {author} {\bibfnamefont {S.}~\bibnamefont {Meenehan}}, \bibinfo {author}
  {\bibfnamefont {J.}~\bibnamefont {Cohen}}, \bibinfo {author} {\bibfnamefont
  {S.}~\bibnamefont {Stobbe}}, \ and\ \bibinfo {author} {\bibfnamefont
  {O.}~\bibnamefont {Painter}},\ }\href {\doibase 10.1364/OE.19.024905}
  {\bibfield  {journal} {\bibinfo  {journal} {Opt. Express}\ }\textbf {\bibinfo
  {volume} {19}},\ \bibinfo {pages} {24905} (\bibinfo {year}
  {2011})}\BibitemShut {NoStop}%
\bibitem [{\citenamefont {Pitanti}\ \emph {et~al.}(2015)\citenamefont
  {Pitanti}, \citenamefont {Fink}, \citenamefont {Safavi-Naeini}, \citenamefont
  {Hill}, \citenamefont {Lei}, \citenamefont {Tredicucci},\ and\ \citenamefont
  {Painter}}]{Pitanti2015}%
  \BibitemOpen
  \bibfield  {author} {\bibinfo {author} {\bibfnamefont {A.}~\bibnamefont
  {Pitanti}}, \bibinfo {author} {\bibfnamefont {J.~M.}\ \bibnamefont {Fink}},
  \bibinfo {author} {\bibfnamefont {A.~H.}\ \bibnamefont {Safavi-Naeini}},
  \bibinfo {author} {\bibfnamefont {J.~T.}\ \bibnamefont {Hill}}, \bibinfo
  {author} {\bibfnamefont {C.~U.}\ \bibnamefont {Lei}}, \bibinfo {author}
  {\bibfnamefont {A.}~\bibnamefont {Tredicucci}}, \ and\ \bibinfo {author}
  {\bibfnamefont {O.}~\bibnamefont {Painter}},\ }\href {\doibase
  10.1364/OE.23.003196} {\bibfield  {journal} {\bibinfo  {journal} {Opt.
  Express}\ }\textbf {\bibinfo {volume} {23}},\ \bibinfo {pages} {3196}
  (\bibinfo {year} {2015})}\BibitemShut {NoStop}%
\bibitem [{\citenamefont {Zhong}\ \emph {et~al.}(2020)\citenamefont {Zhong},
  \citenamefont {Wang}, \citenamefont {Zou}, \citenamefont {Zhang},
  \citenamefont {Han}, \citenamefont {Fu}, \citenamefont {Xu}, \citenamefont
  {Shankar}, \citenamefont {Devoret}, \citenamefont {Tang},\ and\ \citenamefont
  {Jiang}}]{Zhong19}%
  \BibitemOpen
  \bibfield  {author} {\bibinfo {author} {\bibfnamefont {C.}~\bibnamefont
  {Zhong}}, \bibinfo {author} {\bibfnamefont {Z.}~\bibnamefont {Wang}},
  \bibinfo {author} {\bibfnamefont {C.}~\bibnamefont {Zou}}, \bibinfo {author}
  {\bibfnamefont {M.}~\bibnamefont {Zhang}}, \bibinfo {author} {\bibfnamefont
  {X.}~\bibnamefont {Han}}, \bibinfo {author} {\bibfnamefont {W.}~\bibnamefont
  {Fu}}, \bibinfo {author} {\bibfnamefont {M.}~\bibnamefont {Xu}}, \bibinfo
  {author} {\bibfnamefont {S.}~\bibnamefont {Shankar}}, \bibinfo {author}
  {\bibfnamefont {M.~H.}\ \bibnamefont {Devoret}}, \bibinfo {author}
  {\bibfnamefont {H.~X.}\ \bibnamefont {Tang}}, \ and\ \bibinfo {author}
  {\bibfnamefont {L.}~\bibnamefont {Jiang}},\ }\href {\doibase
  10.1103/PhysRevLett.124.010511} {\bibfield  {journal} {\bibinfo  {journal}
  {Phys. Rev. Lett.}\ }\textbf {\bibinfo {volume} {124}},\ \bibinfo {pages}
  {010511} (\bibinfo {year} {2020})}\BibitemShut {NoStop}%
\bibitem [{\citenamefont {Barzanjeh}\ \emph {et~al.}(2012)\citenamefont
  {Barzanjeh}, \citenamefont {Abdi}, \citenamefont {Milburn}, \citenamefont
  {Tombesi},\ and\ \citenamefont {Vitali}}]{Barzanjeh2012}%
  \BibitemOpen
  \bibfield  {author} {\bibinfo {author} {\bibfnamefont {S.}~\bibnamefont
  {Barzanjeh}}, \bibinfo {author} {\bibfnamefont {M.}~\bibnamefont {Abdi}},
  \bibinfo {author} {\bibfnamefont {G.~J.}\ \bibnamefont {Milburn}}, \bibinfo
  {author} {\bibfnamefont {P.}~\bibnamefont {Tombesi}}, \ and\ \bibinfo
  {author} {\bibfnamefont {D.}~\bibnamefont {Vitali}},\ }\href {\doibase
  10.1103/PhysRevLett.109.130503} {\bibfield  {journal} {\bibinfo  {journal}
  {Phys. Rev. Lett.}\ }\textbf {\bibinfo {volume} {109}},\ \bibinfo {pages}
  {130503} (\bibinfo {year} {2012})}\BibitemShut {NoStop}%
\bibitem [{\citenamefont {Rueda}\ \emph {et~al.}(2019)\citenamefont {Rueda},
  \citenamefont {Hease}, \citenamefont {Barzanjeh},\ and\ \citenamefont
  {Fink}}]{rueda2019}%
  \BibitemOpen
  \bibfield  {author} {\bibinfo {author} {\bibfnamefont {A.}~\bibnamefont
  {Rueda}}, \bibinfo {author} {\bibfnamefont {W.}~\bibnamefont {Hease}},
  \bibinfo {author} {\bibfnamefont {S.}~\bibnamefont {Barzanjeh}}, \ and\
  \bibinfo {author} {\bibfnamefont {J.~M.}\ \bibnamefont {Fink}},\ }\href@noop
  {} {\bibfield  {journal} {\bibinfo  {journal} {npj Quantum Information}\
  }\textbf {\bibinfo {volume} {5}},\ \bibinfo {pages} {1} (\bibinfo {year}
  {2019})}\BibitemShut {NoStop}%
\bibitem [{\citenamefont {Rakhubovsky}\ \emph {et~al.}(2016)\citenamefont
  {Rakhubovsky}, \citenamefont {Vostrosablin},\ and\ \citenamefont
  {Filip}}]{Rakhubovsky16}%
  \BibitemOpen
  \bibfield  {author} {\bibinfo {author} {\bibfnamefont {A.~A.}\ \bibnamefont
  {Rakhubovsky}}, \bibinfo {author} {\bibfnamefont {N.}~\bibnamefont
  {Vostrosablin}}, \ and\ \bibinfo {author} {\bibfnamefont {R.}~\bibnamefont
  {Filip}},\ }\href@noop {} {\bibfield  {journal} {\bibinfo  {journal}
  {Physical Review A}\ }\textbf {\bibinfo {volume} {93}},\ \bibinfo {pages}
  {033813} (\bibinfo {year} {2016})}\BibitemShut {NoStop}%
\bibitem [{\citenamefont {Zhang}\ \emph {et~al.}(2018)\citenamefont {Zhang},
  \citenamefont {Zou},\ and\ \citenamefont {Jiang}}]{Zhang18}%
  \BibitemOpen
  \bibfield  {author} {\bibinfo {author} {\bibfnamefont {M.}~\bibnamefont
  {Zhang}}, \bibinfo {author} {\bibfnamefont {C.-L.}\ \bibnamefont {Zou}}, \
  and\ \bibinfo {author} {\bibfnamefont {L.}~\bibnamefont {Jiang}},\
  }\href@noop {} {\bibfield  {journal} {\bibinfo  {journal} {Physical review
  letters}\ }\textbf {\bibinfo {volume} {120}},\ \bibinfo {pages} {020502}
  (\bibinfo {year} {2018})}\BibitemShut {NoStop}%
\bibitem [{\citenamefont {Lau}\ and\ \citenamefont {Clerk}(2019)}]{Hoi19}%
  \BibitemOpen
  \bibfield  {author} {\bibinfo {author} {\bibfnamefont {H.-K.}\ \bibnamefont
  {Lau}}\ and\ \bibinfo {author} {\bibfnamefont {A.~A.}\ \bibnamefont
  {Clerk}},\ }\href@noop {} {\bibfield  {journal} {\bibinfo  {journal} {npj
  Quantum Information}\ }\textbf {\bibinfo {volume} {5}},\ \bibinfo {pages}
  {31} (\bibinfo {year} {2019})}\BibitemShut {NoStop}%
\bibitem [{\citenamefont {Higginbotham}\ \emph {et~al.}()\citenamefont
  {Higginbotham}, \citenamefont {Burns}, \citenamefont {Urmey}, \citenamefont
  {Peterson}, \citenamefont {Kampel}, \citenamefont {Brubaker}, \citenamefont
  {Smith}, \citenamefont {Lehnert},\ and\ \citenamefont
  {Regal}}]{Higginbotham2018}%
  \BibitemOpen
  \bibfield  {author} {\bibinfo {author} {\bibfnamefont {A.~P.}\ \bibnamefont
  {Higginbotham}}, \bibinfo {author} {\bibfnamefont {P.~S.}\ \bibnamefont
  {Burns}}, \bibinfo {author} {\bibfnamefont {M.~D.}\ \bibnamefont {Urmey}},
  \bibinfo {author} {\bibfnamefont {R.~W.}\ \bibnamefont {Peterson}}, \bibinfo
  {author} {\bibfnamefont {N.~S.}\ \bibnamefont {Kampel}}, \bibinfo {author}
  {\bibfnamefont {B.~M.}\ \bibnamefont {Brubaker}}, \bibinfo {author}
  {\bibfnamefont {G.}~\bibnamefont {Smith}}, \bibinfo {author} {\bibfnamefont
  {K.~W.}\ \bibnamefont {Lehnert}}, \ and\ \bibinfo {author} {\bibfnamefont
  {C.~A.}\ \bibnamefont {Regal}},\ }\href@noop {} {\ }\bibinfo {note}
  {\href{https://doi.org/10.1038/s41567-018-0210-0}{Nat. Phys. 14, 1038-1042
  (2018)}}\BibitemShut {NoStop}%
\bibitem [{\citenamefont {Vainsencher}\ \emph {et~al.}(2016)\citenamefont
  {Vainsencher}, \citenamefont {Satzinger}, \citenamefont {Peairs},\ and\
  \citenamefont {Cleland}}]{Vainsencher16}%
  \BibitemOpen
  \bibfield  {author} {\bibinfo {author} {\bibfnamefont {A.}~\bibnamefont
  {Vainsencher}}, \bibinfo {author} {\bibfnamefont {K.~J.}\ \bibnamefont
  {Satzinger}}, \bibinfo {author} {\bibfnamefont {G.~A.}\ \bibnamefont
  {Peairs}}, \ and\ \bibinfo {author} {\bibfnamefont {A.~N.}\ \bibnamefont
  {Cleland}},\ }\href {\doibase 10.1063/1.4955408} {\bibfield  {journal}
  {\bibinfo  {journal} {Appl. Phys. Lett.}\ }\textbf {\bibinfo {volume}
  {109}},\ \bibinfo {pages} {033107} (\bibinfo {year} {2016})}\BibitemShut
  {NoStop}%
\bibitem [{\citenamefont {Han}\ \emph {et~al.}(2014)\citenamefont {Han},
  \citenamefont {Xiong}, \citenamefont {Fong}, \citenamefont {Zhang},\ and\
  \citenamefont {Tang}}]{Han2014}%
  \BibitemOpen
  \bibfield  {author} {\bibinfo {author} {\bibfnamefont {X.}~\bibnamefont
  {Han}}, \bibinfo {author} {\bibfnamefont {C.}~\bibnamefont {Xiong}}, \bibinfo
  {author} {\bibfnamefont {K.~Y.}\ \bibnamefont {Fong}}, \bibinfo {author}
  {\bibfnamefont {X.}~\bibnamefont {Zhang}}, \ and\ \bibinfo {author}
  {\bibfnamefont {H.~X.}\ \bibnamefont {Tang}},\ }\href {\doibase
  10.1088/1367-2630/16/6/063060} {\bibfield  {journal} {\bibinfo  {journal}
  {New J. Phys.}\ }\textbf {\bibinfo {volume} {16}},\ \bibinfo {pages} {063060}
  (\bibinfo {year} {2014})}\BibitemShut {NoStop}%
\bibitem [{\citenamefont {Xu}\ \emph {et~al.}(2019)\citenamefont {Xu},
  \citenamefont {Han}, \citenamefont {Zou}, \citenamefont {Fu}, \citenamefont
  {Xu}, \citenamefont {Zhong}, \citenamefont {Jiang},\ and\ \citenamefont
  {Tang}}]{Mingrui2019}%
  \BibitemOpen
  \bibfield  {author} {\bibinfo {author} {\bibfnamefont {M.}~\bibnamefont
  {Xu}}, \bibinfo {author} {\bibfnamefont {X.}~\bibnamefont {Han}}, \bibinfo
  {author} {\bibfnamefont {C.-L.}\ \bibnamefont {Zou}}, \bibinfo {author}
  {\bibfnamefont {W.}~\bibnamefont {Fu}}, \bibinfo {author} {\bibfnamefont
  {Y.}~\bibnamefont {Xu}}, \bibinfo {author} {\bibfnamefont {C.}~\bibnamefont
  {Zhong}}, \bibinfo {author} {\bibfnamefont {L.}~\bibnamefont {Jiang}}, \ and\
  \bibinfo {author} {\bibfnamefont {H.~X.}\ \bibnamefont {Tang}},\ }\href@noop
  {} {\bibfield  {journal} {\bibinfo  {journal} {arXiv preprint
  arXiv:1910.01203}\ } (\bibinfo {year} {2019})}\BibitemShut {NoStop}%
\bibitem [{\citenamefont {Zou}\ \emph {et~al.}(2016)\citenamefont {Zou},
  \citenamefont {Han}, \citenamefont {Jiang},\ and\ \citenamefont
  {Tang}}]{Zou2016}%
  \BibitemOpen
  \bibfield  {author} {\bibinfo {author} {\bibfnamefont {C.-L.}\ \bibnamefont
  {Zou}}, \bibinfo {author} {\bibfnamefont {X.}~\bibnamefont {Han}}, \bibinfo
  {author} {\bibfnamefont {L.}~\bibnamefont {Jiang}}, \ and\ \bibinfo {author}
  {\bibfnamefont {H.~X.}\ \bibnamefont {Tang}},\ }\href {\doibase
  10.1103/PhysRevA.94.013812} {\bibfield  {journal} {\bibinfo  {journal} {Phys.
  Rev. A}\ }\textbf {\bibinfo {volume} {94}},\ \bibinfo {pages} {013812}
  (\bibinfo {year} {2016})}\BibitemShut {NoStop}%
\bibitem [{\citenamefont {Moiseyev}(2011)}]{Moiseyev11}%
  \BibitemOpen
  \bibfield  {author} {\bibinfo {author} {\bibfnamefont {N.}~\bibnamefont
  {Moiseyev}},\ }\href@noop {} {\emph {\bibinfo {title} {Non-Hermitian quantum
  mechanics}}}\ (\bibinfo  {publisher} {Cambridge University Press},\ \bibinfo
  {year} {2011})\BibitemShut {NoStop}%
\bibitem [{\citenamefont {Clerk}\ \emph {et~al.}(2010)\citenamefont {Clerk},
  \citenamefont {Devoret}, \citenamefont {Girvin}, \citenamefont {Marquardt},\
  and\ \citenamefont {Schoelkopf}}]{Clerk10}%
  \BibitemOpen
  \bibfield  {author} {\bibinfo {author} {\bibfnamefont {A.~A.}\ \bibnamefont
  {Clerk}}, \bibinfo {author} {\bibfnamefont {M.~H.}\ \bibnamefont {Devoret}},
  \bibinfo {author} {\bibfnamefont {S.~M.}\ \bibnamefont {Girvin}}, \bibinfo
  {author} {\bibfnamefont {F.}~\bibnamefont {Marquardt}}, \ and\ \bibinfo
  {author} {\bibfnamefont {R.~J.}\ \bibnamefont {Schoelkopf}},\ }\href
  {\doibase 10.1103/RevModPhys.82.1155} {\bibfield  {journal} {\bibinfo
  {journal} {Rev. Mod. Phys.}\ }\textbf {\bibinfo {volume} {82}},\ \bibinfo
  {pages} {1155} (\bibinfo {year} {2010})}\BibitemShut {NoStop}%
\bibitem [{\citenamefont {Weedbrook}\ \emph {et~al.}(2012)\citenamefont
  {Weedbrook}, \citenamefont {Pirandola}, \citenamefont {Garc\'{\i}a-Patr\'on},
  \citenamefont {Cerf}, \citenamefont {Ralph}, \citenamefont {Shapiro},\ and\
  \citenamefont {Lloyd}}]{Weedbrook2012}%
  \BibitemOpen
  \bibfield  {author} {\bibinfo {author} {\bibfnamefont {C.}~\bibnamefont
  {Weedbrook}}, \bibinfo {author} {\bibfnamefont {S.}~\bibnamefont
  {Pirandola}}, \bibinfo {author} {\bibfnamefont {R.}~\bibnamefont
  {Garc\'{\i}a-Patr\'on}}, \bibinfo {author} {\bibfnamefont {N.~J.}\
  \bibnamefont {Cerf}}, \bibinfo {author} {\bibfnamefont {T.~C.}\ \bibnamefont
  {Ralph}}, \bibinfo {author} {\bibfnamefont {J.~H.}\ \bibnamefont {Shapiro}},
  \ and\ \bibinfo {author} {\bibfnamefont {S.}~\bibnamefont {Lloyd}},\ }\href
  {\doibase 10.1103/RevModPhys.84.621} {\bibfield  {journal} {\bibinfo
  {journal} {Rev. Mod. Phys.}\ }\textbf {\bibinfo {volume} {84}},\ \bibinfo
  {pages} {621} (\bibinfo {year} {2012})}\BibitemShut {NoStop}%
\bibitem [{\citenamefont {De~Gosson}(2006)}]{Gosson06}%
  \BibitemOpen
  \bibfield  {author} {\bibinfo {author} {\bibfnamefont {M.~A.}\ \bibnamefont
  {De~Gosson}},\ }\href@noop {} {\emph {\bibinfo {title} {Symplectic geometry
  and quantum mechanics}}},\ Vol.\ \bibinfo {volume} {166}\ (\bibinfo
  {publisher} {Springer Science \& Business Media},\ \bibinfo {year}
  {2006})\BibitemShut {NoStop}%
\bibitem [{Non()}]{Nonh}%
  \BibitemOpen
  \href@noop {} {}\bibinfo {note} {In general, the hybridized modes is not
  orthogonal to each other due to the matrix $\textbf{M}$ being
  non-hermitian.}\BibitemShut {Stop}%
\bibitem [{\citenamefont {Bender}\ and\ \citenamefont
  {Boettcher}(1998)}]{Bender98}%
  \BibitemOpen
  \bibfield  {author} {\bibinfo {author} {\bibfnamefont {C.~M.}\ \bibnamefont
  {Bender}}\ and\ \bibinfo {author} {\bibfnamefont {S.}~\bibnamefont
  {Boettcher}},\ }\href {\doibase 10.1103/PhysRevLett.80.5243} {\bibfield
  {journal} {\bibinfo  {journal} {Phys. Rev. Lett.}\ }\textbf {\bibinfo
  {volume} {80}},\ \bibinfo {pages} {5243} (\bibinfo {year}
  {1998})}\BibitemShut {NoStop}%
\bibitem [{\citenamefont {Chen}\ \emph {et~al.}(2017)\citenamefont {Chen},
  \citenamefont {{\"O}zdemir}, \citenamefont {Zhao}, \citenamefont {Wiersig},\
  and\ \citenamefont {Yang}}]{Chen2017}%
  \BibitemOpen
  \bibfield  {author} {\bibinfo {author} {\bibfnamefont {W.}~\bibnamefont
  {Chen}}, \bibinfo {author} {\bibfnamefont {{\c{S}}.~K.}\ \bibnamefont
  {{\"O}zdemir}}, \bibinfo {author} {\bibfnamefont {G.}~\bibnamefont {Zhao}},
  \bibinfo {author} {\bibfnamefont {J.}~\bibnamefont {Wiersig}}, \ and\
  \bibinfo {author} {\bibfnamefont {L.}~\bibnamefont {Yang}},\ }\href@noop {}
  {\bibfield  {journal} {\bibinfo  {journal} {Nature}\ }\textbf {\bibinfo
  {volume} {548}},\ \bibinfo {pages} {192} (\bibinfo {year}
  {2017})}\BibitemShut {NoStop}%
\bibitem [{\citenamefont {Marian}\ and\ \citenamefont
  {Marian}(2008)}]{Marian2008}%
  \BibitemOpen
  \bibfield  {author} {\bibinfo {author} {\bibfnamefont {P.}~\bibnamefont
  {Marian}}\ and\ \bibinfo {author} {\bibfnamefont {T.~A.}\ \bibnamefont
  {Marian}},\ }\href {\doibase 10.1103/PhysRevLett.101.220403} {\bibfield
  {journal} {\bibinfo  {journal} {Phys. Rev. Lett.}\ }\textbf {\bibinfo
  {volume} {101}},\ \bibinfo {pages} {220403} (\bibinfo {year}
  {2008})}\BibitemShut {NoStop}%
\bibitem [{\citenamefont {Tserkis}\ and\ \citenamefont
  {Ralph}(2017)}]{Tserkis2017}%
  \BibitemOpen
  \bibfield  {author} {\bibinfo {author} {\bibfnamefont {S.}~\bibnamefont
  {Tserkis}}\ and\ \bibinfo {author} {\bibfnamefont {T.~C.}\ \bibnamefont
  {Ralph}},\ }\href {\doibase 10.1103/PhysRevA.96.062338} {\bibfield  {journal}
  {\bibinfo  {journal} {Phys. Rev. A}\ }\textbf {\bibinfo {volume} {96}},\
  \bibinfo {pages} {062338} (\bibinfo {year} {2017})}\BibitemShut {NoStop}%
\bibitem [{Fil()}]{Filt}%
  \BibitemOpen
  \href@noop {} {}\bibinfo {note} {A single output frequency corresponds to
  using a filter with infinite small bandwidth.}\BibitemShut {Stop}%
\bibitem [{\citenamefont {DeJesus}\ and\ \citenamefont
  {Kaufman}(1987)}]{DeJesus87}%
  \BibitemOpen
  \bibfield  {author} {\bibinfo {author} {\bibfnamefont {E.~X.}\ \bibnamefont
  {DeJesus}}\ and\ \bibinfo {author} {\bibfnamefont {C.}~\bibnamefont
  {Kaufman}},\ }\href {\doibase 10.1103/PhysRevA.35.5288} {\bibfield  {journal}
  {\bibinfo  {journal} {Phys. Rev. A}\ }\textbf {\bibinfo {volume} {35}},\
  \bibinfo {pages} {5288} (\bibinfo {year} {1987})}\BibitemShut {NoStop}%
\bibitem [{\citenamefont {Wang}\ \emph {et~al.}(2015)\citenamefont {Wang},
  \citenamefont {Chesi},\ and\ \citenamefont {Clerk}}]{Yingdan15}%
  \BibitemOpen
  \bibfield  {author} {\bibinfo {author} {\bibfnamefont {Y.-D.}\ \bibnamefont
  {Wang}}, \bibinfo {author} {\bibfnamefont {S.}~\bibnamefont {Chesi}}, \ and\
  \bibinfo {author} {\bibfnamefont {A.~A.}\ \bibnamefont {Clerk}},\ }\href
  {\doibase 10.1103/PhysRevA.91.013807} {\bibfield  {journal} {\bibinfo
  {journal} {Phys. Rev. A}\ }\textbf {\bibinfo {volume} {91}},\ \bibinfo
  {pages} {013807} (\bibinfo {year} {2015})}\BibitemShut {NoStop}%
\bibitem [{\citenamefont {Tian}(2013)}]{Lin13}%
  \BibitemOpen
  \bibfield  {author} {\bibinfo {author} {\bibfnamefont {L.}~\bibnamefont
  {Tian}},\ }\href {\doibase 10.1103/PhysRevLett.110.233602} {\bibfield
  {journal} {\bibinfo  {journal} {Phys. Rev. Lett.}\ }\textbf {\bibinfo
  {volume} {110}},\ \bibinfo {pages} {233602} (\bibinfo {year}
  {2013})}\BibitemShut {NoStop}%
\bibitem [{int()}]{inter}%
  \BibitemOpen
  \href@noop {} {}\bibinfo {note} {Strictly, the $\hat{a}_1$, $\hat{a}_2$,
  $\hat{B}$ and $\hat{C}$ mode operators here should represent the output
  modes. We left out the subscript ``out" for simplicity.}\BibitemShut {Stop}%
\bibitem [{Nar()}]{Narr}%
  \BibitemOpen
  \href@noop {} {}\bibinfo {note} {The filter should have bandwidth in several
  MHz. This extremely narrow band filter has been demonstrated with rare-earth
  ion doped crystals.}\BibitemShut {Stop}%
\bibitem [{\citenamefont {Beavan}\ \emph {et~al.}(2013)\citenamefont {Beavan},
  \citenamefont {Goldschmidt},\ and\ \citenamefont {Sellars}}]{Sarah13}%
  \BibitemOpen
  \bibfield  {author} {\bibinfo {author} {\bibfnamefont {S.~E.}\ \bibnamefont
  {Beavan}}, \bibinfo {author} {\bibfnamefont {E.~A.}\ \bibnamefont
  {Goldschmidt}}, \ and\ \bibinfo {author} {\bibfnamefont {M.~J.}\ \bibnamefont
  {Sellars}},\ }\href {\doibase 10.1364/JOSAB.30.001173} {\bibfield  {journal}
  {\bibinfo  {journal} {J. Opt. Soc. Am. B}\ }\textbf {\bibinfo {volume}
  {30}},\ \bibinfo {pages} {1173} (\bibinfo {year} {2013})}\BibitemShut
  {NoStop}%
\bibitem [{\citenamefont {Guo}\ \emph {et~al.}(2017)\citenamefont {Guo},
  \citenamefont {Mei},\ and\ \citenamefont {Du}}]{Xianxin17}%
  \BibitemOpen
  \bibfield  {author} {\bibinfo {author} {\bibfnamefont {X.}~\bibnamefont
  {Guo}}, \bibinfo {author} {\bibfnamefont {Y.}~\bibnamefont {Mei}}, \ and\
  \bibinfo {author} {\bibfnamefont {S.}~\bibnamefont {Du}},\ }\href {\doibase
  10.1364/OPTICA.4.000388} {\bibfield  {journal} {\bibinfo  {journal} {Optica}\
  }\textbf {\bibinfo {volume} {4}},\ \bibinfo {pages} {388} (\bibinfo {year}
  {2017})}\BibitemShut {NoStop}%
\bibitem [{\citenamefont {Campagne-Ibarcq}\ \emph {et~al.}(2018)\citenamefont
  {Campagne-Ibarcq}, \citenamefont {Zalys-Geller}, \citenamefont {Narla},
  \citenamefont {Shankar}, \citenamefont {Reinhold}, \citenamefont {Burkhart},
  \citenamefont {Axline}, \citenamefont {Pfaff}, \citenamefont {Frunzio},
  \citenamefont {Schoelkopf},\ and\ \citenamefont
  {Devoret}}]{Campagne-Ibarcq2018}%
  \BibitemOpen
  \bibfield  {author} {\bibinfo {author} {\bibfnamefont {P.}~\bibnamefont
  {Campagne-Ibarcq}}, \bibinfo {author} {\bibfnamefont {E.}~\bibnamefont
  {Zalys-Geller}}, \bibinfo {author} {\bibfnamefont {A.}~\bibnamefont {Narla}},
  \bibinfo {author} {\bibfnamefont {S.}~\bibnamefont {Shankar}}, \bibinfo
  {author} {\bibfnamefont {P.}~\bibnamefont {Reinhold}}, \bibinfo {author}
  {\bibfnamefont {L.}~\bibnamefont {Burkhart}}, \bibinfo {author}
  {\bibfnamefont {C.}~\bibnamefont {Axline}}, \bibinfo {author} {\bibfnamefont
  {W.}~\bibnamefont {Pfaff}}, \bibinfo {author} {\bibfnamefont
  {L.}~\bibnamefont {Frunzio}}, \bibinfo {author} {\bibfnamefont {R.~J.}\
  \bibnamefont {Schoelkopf}}, \ and\ \bibinfo {author} {\bibfnamefont {M.~H.}\
  \bibnamefont {Devoret}},\ }\href {\doibase 10.1103/PhysRevLett.120.200501}
  {\bibfield  {journal} {\bibinfo  {journal} {Phys. Rev. Lett.}\ }\textbf
  {\bibinfo {volume} {120}},\ \bibinfo {pages} {200501} (\bibinfo {year}
  {2018})}\BibitemShut {NoStop}%
\bibitem [{Esw()}]{Eswap}%
  \BibitemOpen
  \href@noop {} {}\bibinfo {note} {The swap realizes
  $\ket{1}_B\ket{g}_{Q_1}\rightarrow\ket{0}_B\ket{e}_{Q_1}$ and
  $\ket{1}_C\ket{g}_{Q_2}\rightarrow\ket{0}_C\ket{e}_{Q_2}$, which gives an
  entangled state between optical photons and the transmon qubits
  $\frac{\sqrt{2}}{2}(\ket{1}_{a_1}\ket{e}_{Q_1}+\ket{1}_{a_2}\ket{e}_{Q_2})$}\BibitemShut
  {NoStop}%
\bibitem [{\citenamefont {Nielsen}\ and\ \citenamefont
  {Chuang}(2002)}]{Nielsen02}%
  \BibitemOpen
  \bibfield  {author} {\bibinfo {author} {\bibfnamefont {M.~A.}\ \bibnamefont
  {Nielsen}}\ and\ \bibinfo {author} {\bibfnamefont {I.}~\bibnamefont
  {Chuang}},\ }\href@noop {} {\emph {\bibinfo {title} {Quantum computation and
  quantum information}}}\ (\bibinfo  {publisher} {AAPT},\ \bibinfo {year}
  {2002})\BibitemShut {NoStop}%
\bibitem [{\citenamefont {Shankar}\ \emph {et~al.}(2013)\citenamefont
  {Shankar}, \citenamefont {Hatridge}, \citenamefont {Leghtas}, \citenamefont
  {Sliwa}, \citenamefont {Narla}, \citenamefont {Vool}, \citenamefont {Girvin},
  \citenamefont {Frunzio}, \citenamefont {Mirrahimi},\ and\ \citenamefont
  {Devoret}}]{Shankar2013}%
  \BibitemOpen
  \bibfield  {author} {\bibinfo {author} {\bibfnamefont {S.}~\bibnamefont
  {Shankar}}, \bibinfo {author} {\bibfnamefont {M.}~\bibnamefont {Hatridge}},
  \bibinfo {author} {\bibfnamefont {Z.}~\bibnamefont {Leghtas}}, \bibinfo
  {author} {\bibfnamefont {K.~M.}\ \bibnamefont {Sliwa}}, \bibinfo {author}
  {\bibfnamefont {A.}~\bibnamefont {Narla}}, \bibinfo {author} {\bibfnamefont
  {U.}~\bibnamefont {Vool}}, \bibinfo {author} {\bibfnamefont {S.~M.}\
  \bibnamefont {Girvin}}, \bibinfo {author} {\bibfnamefont {L.}~\bibnamefont
  {Frunzio}}, \bibinfo {author} {\bibfnamefont {M.}~\bibnamefont {Mirrahimi}},
  \ and\ \bibinfo {author} {\bibfnamefont {M.~H.}\ \bibnamefont {Devoret}},\
  }\href {http://dx.doi.org/10.1038/nature12802} {\bibfield  {journal}
  {\bibinfo  {journal} {Nature}\ }\textbf {\bibinfo {volume} {504}},\ \bibinfo
  {pages} {419} (\bibinfo {year} {2013})}\BibitemShut {NoStop}%
\bibitem [{\citenamefont {Hatridge}\ \emph {et~al.}(2013)\citenamefont
  {Hatridge}, \citenamefont {Shankar}, \citenamefont {Mirrahimi}, \citenamefont
  {Schackert}, \citenamefont {Geerlings}, \citenamefont {Brecht}, \citenamefont
  {Sliwa}, \citenamefont {Abdo}, \citenamefont {Frunzio}, \citenamefont
  {Girvin}, \citenamefont {Schoelkopf},\ and\ \citenamefont
  {Devoret}}]{Hatridge2013}%
  \BibitemOpen
  \bibfield  {author} {\bibinfo {author} {\bibfnamefont {M.}~\bibnamefont
  {Hatridge}}, \bibinfo {author} {\bibfnamefont {S.}~\bibnamefont {Shankar}},
  \bibinfo {author} {\bibfnamefont {M.}~\bibnamefont {Mirrahimi}}, \bibinfo
  {author} {\bibfnamefont {F.}~\bibnamefont {Schackert}}, \bibinfo {author}
  {\bibfnamefont {K.}~\bibnamefont {Geerlings}}, \bibinfo {author}
  {\bibfnamefont {T.}~\bibnamefont {Brecht}}, \bibinfo {author} {\bibfnamefont
  {K.~M.}\ \bibnamefont {Sliwa}}, \bibinfo {author} {\bibfnamefont
  {B.}~\bibnamefont {Abdo}}, \bibinfo {author} {\bibfnamefont {L.}~\bibnamefont
  {Frunzio}}, \bibinfo {author} {\bibfnamefont {S.~M.}\ \bibnamefont {Girvin}},
  \bibinfo {author} {\bibfnamefont {R.~J.}\ \bibnamefont {Schoelkopf}}, \ and\
  \bibinfo {author} {\bibfnamefont {M.~H.}\ \bibnamefont {Devoret}},\ }\href
  {\doibase 10.1126/science.1226897} {\bibfield  {journal} {\bibinfo  {journal}
  {Science}\ }\textbf {\bibinfo {volume} {339}},\ \bibinfo {pages} {178}
  (\bibinfo {year} {2013})}\BibitemShut {NoStop}%
\bibitem [{\citenamefont {Clauser}\ \emph {et~al.}(1969)\citenamefont
  {Clauser}, \citenamefont {Horne}, \citenamefont {Shimony},\ and\
  \citenamefont {Holt}}]{Clauser1969}%
  \BibitemOpen
  \bibfield  {author} {\bibinfo {author} {\bibfnamefont {J.~F.}\ \bibnamefont
  {Clauser}}, \bibinfo {author} {\bibfnamefont {M.~A.}\ \bibnamefont {Horne}},
  \bibinfo {author} {\bibfnamefont {A.}~\bibnamefont {Shimony}}, \ and\
  \bibinfo {author} {\bibfnamefont {R.~A.}\ \bibnamefont {Holt}},\ }\href
  {\doibase 10.1103/PhysRevLett.23.880} {\bibfield  {journal} {\bibinfo
  {journal} {Phys. Rev. Lett.}\ }\textbf {\bibinfo {volume} {23}},\ \bibinfo
  {pages} {880} (\bibinfo {year} {1969})}\BibitemShut {NoStop}%
\bibitem [{\citenamefont {Glauber}(1963{\natexlab{a}})}]{Glauber636}%
  \BibitemOpen
  \bibfield  {author} {\bibinfo {author} {\bibfnamefont {R.~J.}\ \bibnamefont
  {Glauber}},\ }\href {\doibase 10.1103/PhysRev.130.2529} {\bibfield  {journal}
  {\bibinfo  {journal} {Phys. Rev.}\ }\textbf {\bibinfo {volume} {130}},\
  \bibinfo {pages} {2529} (\bibinfo {year} {1963}{\natexlab{a}})}\BibitemShut
  {NoStop}%
\bibitem [{\citenamefont {Glauber}(1963{\natexlab{b}})}]{Glauber639}%
  \BibitemOpen
  \bibfield  {author} {\bibinfo {author} {\bibfnamefont {R.~J.}\ \bibnamefont
  {Glauber}},\ }\href {\doibase 10.1103/PhysRev.131.2766} {\bibfield  {journal}
  {\bibinfo  {journal} {Phys. Rev.}\ }\textbf {\bibinfo {volume} {131}},\
  \bibinfo {pages} {2766} (\bibinfo {year} {1963}{\natexlab{b}})}\BibitemShut
  {NoStop}%
\bibitem [{\citenamefont {Shapiro}\ and\ \citenamefont
  {Sun}(1994)}]{Shapiro94}%
  \BibitemOpen
  \bibfield  {author} {\bibinfo {author} {\bibfnamefont {J.~H.}\ \bibnamefont
  {Shapiro}}\ and\ \bibinfo {author} {\bibfnamefont {K.-X.}\ \bibnamefont
  {Sun}},\ }\href
  {https://www.osapublishing.org/josab/abstract.cfm?uri=josab-11-6-1130}
  {\bibfield  {journal} {\bibinfo  {journal} {JOSA B}\ }\textbf {\bibinfo
  {volume} {11}},\ \bibinfo {pages} {1130} (\bibinfo {year}
  {1994})}\BibitemShut {NoStop}%
\bibitem [{\citenamefont {Bennett}\ \emph {et~al.}(1993)\citenamefont
  {Bennett}, \citenamefont {Brassard}, \citenamefont {Cr\'epeau}, \citenamefont
  {Jozsa}, \citenamefont {Peres},\ and\ \citenamefont {Wootters}}]{Bennett93}%
  \BibitemOpen
  \bibfield  {author} {\bibinfo {author} {\bibfnamefont {C.~H.}\ \bibnamefont
  {Bennett}}, \bibinfo {author} {\bibfnamefont {G.}~\bibnamefont {Brassard}},
  \bibinfo {author} {\bibfnamefont {C.}~\bibnamefont {Cr\'epeau}}, \bibinfo
  {author} {\bibfnamefont {R.}~\bibnamefont {Jozsa}}, \bibinfo {author}
  {\bibfnamefont {A.}~\bibnamefont {Peres}}, \ and\ \bibinfo {author}
  {\bibfnamefont {W.~K.}\ \bibnamefont {Wootters}},\ }\href {\doibase
  10.1103/PhysRevLett.70.1895} {\bibfield  {journal} {\bibinfo  {journal}
  {Phys. Rev. Lett.}\ }\textbf {\bibinfo {volume} {70}},\ \bibinfo {pages}
  {1895} (\bibinfo {year} {1993})}\BibitemShut {NoStop}%
\bibitem [{\citenamefont {Bouwmeester}\ \emph {et~al.}(1997)\citenamefont
  {Bouwmeester}, \citenamefont {Pan}, \citenamefont {Mattle}, \citenamefont
  {Eibl}, \citenamefont {Weinfurter},\ and\ \citenamefont
  {Zeilinger}}]{Bouwmeester97}%
  \BibitemOpen
  \bibfield  {author} {\bibinfo {author} {\bibfnamefont {D.}~\bibnamefont
  {Bouwmeester}}, \bibinfo {author} {\bibfnamefont {J.-W.}\ \bibnamefont
  {Pan}}, \bibinfo {author} {\bibfnamefont {K.}~\bibnamefont {Mattle}},
  \bibinfo {author} {\bibfnamefont {M.}~\bibnamefont {Eibl}}, \bibinfo {author}
  {\bibfnamefont {H.}~\bibnamefont {Weinfurter}}, \ and\ \bibinfo {author}
  {\bibfnamefont {A.}~\bibnamefont {Zeilinger}},\ }\href
  {https://www.nature.com/articles/37539} {\bibfield  {journal} {\bibinfo
  {journal} {Nature}\ }\textbf {\bibinfo {volume} {390}},\ \bibinfo {pages}
  {575} (\bibinfo {year} {1997})}\BibitemShut {NoStop}%
\bibitem [{\citenamefont {Duan}\ \emph {et~al.}(2001)\citenamefont {Duan},
  \citenamefont {Lukin}, \citenamefont {Cirac},\ and\ \citenamefont
  {Zoller}}]{duan2001}%
  \BibitemOpen
  \bibfield  {author} {\bibinfo {author} {\bibfnamefont {L.-M.}\ \bibnamefont
  {Duan}}, \bibinfo {author} {\bibfnamefont {M.~D.}\ \bibnamefont {Lukin}},
  \bibinfo {author} {\bibfnamefont {J.~I.}\ \bibnamefont {Cirac}}, \ and\
  \bibinfo {author} {\bibfnamefont {P.}~\bibnamefont {Zoller}},\ }\href
  {http://dx.doi.org/10.1038/35106500} {\bibfield  {journal} {\bibinfo
  {journal} {Nature}\ }\textbf {\bibinfo {volume} {414}},\ \bibinfo {pages}
  {413} (\bibinfo {year} {2001})}\BibitemShut {NoStop}%
\bibitem [{\citenamefont {Ferraro}\ \emph {et~al.}()\citenamefont {Ferraro},
  \citenamefont {Olivares},\ and\ \citenamefont {Paris}}]{Ferraro2005}%
  \BibitemOpen
  \bibfield  {author} {\bibinfo {author} {\bibfnamefont {A.}~\bibnamefont
  {Ferraro}}, \bibinfo {author} {\bibfnamefont {S.}~\bibnamefont {Olivares}}, \
  and\ \bibinfo {author} {\bibfnamefont {M.~G.}\ \bibnamefont {Paris}},\ }\href
  {https://arxiv.org/abs/quant-ph/0503237} {\ }\Eprint
  {http://arxiv.org/abs/quant-ph/0503237} {arXiv:quant-ph/0503237} \BibitemShut
  {NoStop}%
\end{thebibliography}%

\end{document}